\documentclass[manuscript,screen]{acmart}

\usepackage{packages}

\AtBeginDocument{%
  }

\setcopyright{rightsretained}
\acmJournal{TOSEM}
\acmYear{2024} \acmVolume{1} \acmNumber{1} \acmArticle{1} \acmMonth{1}\acmDOI{10.1145/3678172}

\begin{document}

\title{Investigating the Role of Cultural Values in Adopting Large Language Models for Software Engineering}

\author{Stefano Lambiase}
\email{slambiase@unisa.it}
\orcid{0000-0002-9933-6203}
\affiliation{
  \institution{University of Salerno}
  \city{Fisciano}
  \state{(SA)}
  \country{Italy}
}

\author{Gemma Catolino}
\email{gcatolino@unisa.it}
\orcid{0000-0002-4689-3401}
\affiliation{
  \institution{University of Salerno}
  \city{Fisciano}
  \state{(SA)}
  \country{Italy}
}

\author{Fabio Palomba}
\email{fpalomba@unisa.it}
\orcid{0000-0001-9337-5116}
\affiliation{
  \institution{University of Salerno}
  \city{Fisciano}
  \state{(SA)}
  \country{Italy}
}

\author{Filomena Ferrucci}
\email{fferrucci@unisa.it}
\orcid{0000-0002-0975-8972}
\affiliation{
  \institution{University of Salerno}
  \city{Fisciano}
  \state{(SA)}
  \country{Italy}
}

\author{Daniel Russo}
\email{daniel.russo@cs.aau.dk}
\orcid{0000-0001-7253-101X}
\affiliation{
  \institution{Aalborg University Copenhagen}
  \city{Copenhagen}
  \state{Hovedstaden}
  \country{Denmark}
}

\renewcommand{\shortauthors}{Lambiase et al.}

\begin{abstract}
As a socio-technical activity, software development involves the close interconnection of people and technology. The integration of Large Language Models (LLMs) into this process exemplifies the socio-technical nature of software development. Although LLMs influence the development process, software development remains fundamentally human-centric, necessitating an investigation of the human factors in this adoption. Thus, with this study we explore the factors influencing the adoption of LLMs in software development, focusing on the role of professionals' cultural values. Guided by the Unified Theory of Acceptance and Use of Technology (UTAUT2) and Hofstede's cultural dimensions, we hypothesized that cultural values moderate the relationships within the UTAUT2 framework. Using Partial Least Squares-Structural Equation Modelling and data from 188 software engineers, we found that habit and performance expectancy are the primary drivers of LLM adoption, while cultural values do not significantly moderate this process. These findings suggest that, by highlighting how LLMs can boost performance and efficiency, organizations can encourage their use, no matter the cultural differences. Practical steps include offering training programs to demonstrate LLM benefits, creating a supportive environment for regular use, and continuously tracking and sharing performance improvements from using LLMs.

\end{abstract}



\maketitle

\section{Introduction}

\begin{quote}
“Technology is nothing. What's important is that you have faith in people, that they're basically good and smart -- and if you give them tools, they'll do wonderful things with them. Tools are just tools. They either work, or they don't work.”

\begin{flushright}
\textit{--- Steve Jobs}
\end{flushright}
\end{quote}
\vspace{2mm}

Generative AI has become increasingly pervasive in individuals' daily lives. In particular, the advent of Large Language Models (LLMs) has significantly disrupted various activities, notably transforming software development. In fact, a large body of knowledge is starting to grow investigating the use of LLMs for supporting software development and engineering activities. For such a reason, the research community has taken a keen interest in this impact, reporting that professionals are already leveraging LLMs in multiple ways, including generating source code, creating entire software modules, providing decision support, and identifying problems to enhance maintenance activities~\cite{russo2024_navigating, kumar2024code, liu2024empirical, draxler2023gender, agossah2023llm, khojah2024beyond}.

Despite the integration of LLMs into the software development process, this field remains (still) inherently socio-technical. It involves collaboration among individuals with diverse cultural backgrounds, skill sets, and ways of thinking, all working together and with technology to achieve a common objective~\cite{MyticalManMonth, Social_Theory, hoda2021_STGT, naqvi2023_socio_technical, cataldo2008_socio_technical_congruence}. At present, practitioners view LLMs as companions in their daily work activities; they leverage the capabilities of LLMs in various ways and to different extents, with these variations being influenced by a wide range of personal preferences and social factors~\cite{russo2024_navigating, kumar2024code, liu2024empirical}.
Given the transformative potential of LLMs in software engineering, it is crucial to better understand their implications on both the technical and social aspects of development. As LLMs become more integrated into the workflow, they have the potential to reshape not only the tools and practices of software engineering but also the dynamics of team collaboration, decision-making processes, and the overall role of engineers. Understanding the factors that led the adoption of LLMs is essential for optimizing their use and mitigating any potential drawbacks.

Among the human factors influencing individual choices, cultural values stand out as a significant determinant~\cite{lai_2016_culture_and_teaching_both_approaches, baptista_2015_culture_and_mobile_banking, nistor_2014_culture_and_learning_tools_computers, tamilmani_2021_UTAUT2_LR} whose potential influence on LLMs adoption remains unexplored. Culture encompasses shared motives, values, beliefs, identities, and interpretations of significant events that arise from everyday experiences within groups and underlie human attitudes and behaviors~\cite{hofstede_1980_cultural_model_1, markus2014_culture_Definition, house2004_GLOBE, house2013_GLOBE, chhokar2007_GLOBE}. Researchers in both software engineering and management have demonstrated that the values deriving from the cultural background—from now on, \textit{cultural values}—can influence a wide range of variables, including individual personality traits and the choice to adopt a specific technology~\cite{lai_2016_culture_and_teaching_both_approaches, baptista_2015_culture_and_mobile_banking, nistor_2014_culture_and_learning_tools_computers, tamilmani_2021_UTAUT2_LR}. As Im, Hong, and Kang~\cite{im2011international} claim, technology adoption is as much a cultural issue as a rational decision-making process. Given that the adoption of LLMs, like other technologies, is influenced by both technical and social aspects, it is reasonable to assume that cultural background plays a role in this process. Thus, exploring how cultural factors impact the acceptance and use of LLMs could provide valuable insights, ultimately facilitating their seamless integration into the software development industry. Such a smooth integration ensures that LLMs are effectively leveraged, enhancing productivity, fostering innovation, and minimizing resistance or challenges that could arise from cultural misalignment or management misunderstandings.

Starting from what we mentioned above, the present research started from the following research question:

\steResearchQuestionBox{\faBullseye \hspace{0.05cm} \textbf{Research Question.} \textit{What is the role of individuals' cultural values in adopting Large Language Models for supporting software engineering activities?}}

To address the research question, our study was divided into two key objectives. First, we explored the individual factors associated with technology adoption as identified in the literature, using the \textit{Unified Theory of Acceptance and Use of Technology} (UTAUT2)~\cite{Venkatesh_2012_UTAUT2} as a guiding framework. Second, we examined the influence of individual cultural values on these factors and on technology adoption itself, operationalizing Hofstede's cultural dimensions to conceptualize participants' cultural backgrounds~\cite{hofstede_1980_cultural_model_1, hofstede_2011_cultural_model_2}.

The UTAUT2 aims to explain users' intentions and behaviors related to technology acceptance. It includes constructs such as performance expectancy, effort expectancy, social influence, facilitating conditions, hedonic motivation, price value, and habit. On the other side, Hofstede's cultural values framework identifies six dimensions of culture: power distance, individualism vs. collectivism, masculinity vs. femininity, uncertainty avoidance, long-term orientation vs. short-term normative orientation, and indulgence vs. restraint. These dimensions help in understanding how cultural differences impact various behaviors and attitudes. Furthermore, it is important to emphasize that we do not rely on the cultural dimensions at national level—i.e., measuring the cultural values of an individual relying on their birth place—but at individual level, measured through the largely used Yoo et al. CVSCALE~\cite{yoo_2011_CVSCALE_measuring_culture}; this is the correct approach when studying a phenomenon related to the individual behavior.

From the literature, we derived a theoretical model incorporating the two concepts mentioned above. In such model, we integrated the constructs associated with Hofstede's cultural values as \textit{moderators} in the UTAUT2 model\footnote{A moderator effect occurs when the relationship between two variables changes depending on the level of a third variable~\cite{hair_2014_PLS}, which in this case are the cultural values. For example, in cultures with high collectivism, the positive impact of Performance Expectancy on Behavioral Intention might be stronger because individuals in these cultures place more value on technologies that benefit the group. Conversely, in cultures with low collectivism, the influence of Performance Expectancy on Behavioral Intention might be weaker, as individual benefits are prioritized over group benefits.}. The choice to use cultural values as moderators have been guided by similar literature on the field of cultural influence in technology adoption~\cite{lai_2016_culture_and_teaching_both_approaches, tamilmani_2021_UTAUT2_LR, baptista_2015_culture_and_mobile_banking, nistor_2014_culture_and_learning_tools_computers}.

We tested our theory using \textit{Partial Least Squares-Structural Equation Modelling} (PLS-SEM)~\cite{hair_2014_PLS, russo2021_pls_SLR}. PLS-SEM is a statistical technique for analyzing complex relationships between observed and latent variables, combining features of factor analysis and multiple regression to simultaneously examine multiple dependent and independent variables. It is particularly valuable in exploratory research and for models involving complex constructs, such as cultural values. Data collection was conducted through two questionnaires: one for pre-screening participants and another for measuring the construct. These questionnaires aimed to capture the adoption behavior of LLMs and the cultural background of 188 software engineers.



We argue that gaining a deeper understanding of the adoption of LLMs in software engineering could offer numerous actionable insights and lay the groundwork for broader research opportunities.
\begin{itemize}
    \item First, while the interrelationship between culture and technology is a recognized research topic in many disciplines~\cite{lai_2016_culture_and_teaching_both_approaches, tamilmani_2021_UTAUT2_LR, baptista_2015_culture_and_mobile_banking, nistor_2014_culture_and_learning_tools_computers, srite2006_national_culture_TAM, hoehle_2015_national_culture_TAM}, there is a lack of studies in software engineering. This gap in knowledge could lead researchers to incorrect conclusions, resulting in incomplete transfer of knowledge from research to practice. We argue that our work could serve as a milestone in examining the role of cultural values in the adoption of technology in software engineering. Furthermore, although culture has been shown to be important in several technology adoption processes, to date there is no research work covering its role in the adoption of LLMs in software development teams. 

    \item Second, although some researchers in software engineering have investigated culture and its implications, most have conceptualized culture at the national level. When studying phenomena at the individual level, this approach—often based on the scores provided by various cultural frameworks (such as those of Hofstede~\cite{hofstede_2011_cultural_model_2} or GLOBE~\cite{house2004_GLOBE})—fail to capture the substantial variations in cultural values among individuals within the same country, leading to questionable results~\cite{fang2009_culture_critique, fang2012_culture_critique, ford2003_culture_critique, neumann2024_culture_critique}. To avoid this fallacy, we heeded the call of scholars advocating for a better representation of culture~\cite{srite2006_national_culture_TAM, neumann2024_culture_critique} and used individual espoused cultural values in our study instead of national scores. To the best of our knowledge, this is a novel approach in the field of software engineering research, which still relies heavily on national scores.

    \item Third, combining cultural values with UTAUT2 constructs to investigate the adoption of LLMs in software engineering has the potential to provide concrete and actionable results~\cite{lai_2016_culture_and_teaching_both_approaches, tamilmani_2021_UTAUT2_LR, baptista_2015_culture_and_mobile_banking, nistor_2014_culture_and_learning_tools_computers, srite2006_national_culture_TAM, hoehle_2015_national_culture_TAM}. By identifying the key factors influencing the use of this disruptive technology, we aimed to offer a set of actions that organizations can adopt to develop culturally-sensitive approaches to fostering LLMs adoption. For example, organizations could implement tools, procedures, and cross-sectional processes to integrate the technology into their teams' workflows. This careful integration can mitigate ethical concerns and prevent disruptive situations with potentially catastrophic consequences for project outcomes.

    \item Furthermore, our research could benefit other researchers by paving the way for new research agendas regarding generative AI in software development. It could also inspire others to investigate cultural values with greater awareness, leading to new research projects in similar areas.
\end{itemize}

In summary, this research makes the following contributions:

\begin{itemize}
    \item Conducted a statistical evaluation using PLS-SEM to assess the impact of factors from the UTAUT2 model on the intention to use and actual adoption of LLMs for software engineering.
    \item Performed a moderating analysis using PLS-SEM to examine the role of cultural values in the intention to use and adoption of LLMs for software engineering.
    \item Made all tests and data publicly available in an online appendix~\cite{online_appendix} to ensure reliability and promote open-source collaboration.
\end{itemize}

The paper is structured as follows. Section \ref{sec_background} presents the related work. Section \ref{sec_theory} reports the theoretical constructs that lead to the definition of hypothesis that are foundational for the tested theory. In Section \ref{sec_method}, we present the research method and steps adopted to answer our research question. Section \ref{sec_results} reports the results of the statistical analysis, while Section \ref{sec_discussion} presents some reflections and implications of our study. Section \ref{sec_conclusion} closes our research paper with some conclusions and our future research agenda.



\section{Related Work}
\label{sec_background}

Large Language Models (LLMs) are advanced artificial intelligence systems designed to understand and generate human-like text by learning statistical patterns from vast datasets. Originating from the field of natural language processing, LLMs have demonstrated significant potential across various domains, including software development~\cite{brown2020language}. In the context of software development, LLMs are particularly useful for tasks such as code translation, code generation, and code description, providing valuable assistance even when errors occur by offering useful hints for solutions~\cite{liu2024empirical, kumar2024code, khojah2024beyond}. These models, such as OpenAI's GPT-3.5-Turbo and GPT-4-Turbo, as well as open-source variants like Meta's Codellama and Deepseek's Deepseek Coder, have been explored for their capabilities in generating tests to validate and verify compiler implementations, showcasing their impressive code generation abilities. Overall, LLMs originate from the need to model and generate human language, and their application in software development and other fields underscores their broad utility and evolving capabilities.

The research community is beginning to explore the adoption of LLMs and the factors influencing this process more deeply. In the following section, we outline current research trends related to LLM adoption, which provide valuable insights for understanding our work and its position within the state of the art, as the research gap we are addressing.

\subsection{Use of Large Language Models in Software Development}

The application of LLMs in software development is revolutionizing the field, providing innovative solutions that enhance both efficiency and quality. Recent research highlights the diverse ways LLMs are being leveraged across various software engineering tasks, from code clone detection and summarization to safety-critical code generation and problem-solving guidance. One significant application of LLMs is in code clone detection, code summarization, and program comprehension, as demonstrated by Kumar and Chimalakonda through their effective use of federated learning (FedLLM)~\cite{kumar2024code}. In the realm of safety-critical software development, LLMs like GPT-4 are being employed to generate code for high-stakes industries, with Liu et al. introducing the novel prompt engineering method, Prompt-FDC, which enhances code completeness, comment rate, and overall quality~\cite{liu2024empirical}. Beyond these specific tasks, LLMs are also transforming how professional software engineers approach their day-to-day work. Khojah et al. explore the use of ChatGPT by engineers, revealing its utility for guidance, learning, and solving complex engineering problems~\cite{khojah2024beyond}. Given that this topic is still in its early stages~\cite{Gartner2023_AI}, many research papers are currently under review. These upcoming studies promise to provide a deeper and more comprehensive understanding of the use of LLMs in software engineering, further elucidating their potential and addressing existing challenges. The ongoing research will undoubtedly contribute to refining these technologies and broadening their application, paving the way for even more significant advancements in the field.

\subsection{Benefits and Challenges of Adopting Large Language Models}

LLMs offer numerous benefits in software development, significantly enhancing productivity and capabilities across various tasks. One of the primary advantages, as reported by Fernandes~\cite{fernandes2023programming}, is their ability to perform code translation, generation, and description, which can be particularly useful during software development activities~\cite{fernandes2023programming}. Ross et al.~\cite{ross2023programmer} integrated LLMs into development tools to assist with translating code between programming languages, generating code from natural language, and autocompleting code, thereby streamlining the coding process and reducing the time developers spend on routine tasks~\cite{ross2023programmer}. Moreover, LLMs can engage in extended, multi-turn discussions with developers, providing a more interactive and conversational approach to coding assistance, which has been shown to improve productivity and uncover additional knowledge and capabilities beyond mere code generation~\cite{ross2023programmer}. In the realm of testing and validation, Munley et al.~\cite{munley2024llm4vv} demonstrated LLMs utility by automatically generating tests to validate compiler implementations, such as those for the OpenACC parallel programming paradigm, with some models like Deepseek-Coder-33b-Instruct and GPT-4-Turbo producing highly effective results~\cite{munley2024llm4vv}. Additionally, LLMs are being leveraged in educational settings to help students generate code and assist instructors in creating learning materials, thus enhancing the learning experience and fostering a more inclusive educational community~\cite{macneil2023implications}. Finally, LLMs have shown promise in improving code security by detecting vulnerabilities and protecting data privacy, outperforming traditional methods in these areas. However, it is important to note that their advanced capabilities also pose potential risks, such as being exploited for user-level attacks due to their human-like reasoning abilities~\cite{yao2024survey}.

\subsection{Factors Influencing the Adoption of Large Language Models}

The adoption of LLMs in professional contexts is shaped by several common factors identified across multiple studies. Perceived usefulness consistently emerges as a critical factor, with users more likely to adopt LLMs if they find them valuable for their tasks, as noted by both Agossah et al.~\cite{agossah2023llm} and Khojah et al.~\cite{khojah2024beyond}. The frequency of use also plays a significant role, as regular interaction with LLMs reinforces their perceived utility, highlighting the importance of sustained engagement. Education and expertise are crucial, with Draxler et al.~\cite{draxler2023gender} emphasizing the need for equitable access to technology-related education to maximize LLM benefits and bridge gaps such as gender disparities. Additionally, the ease of integration into existing workflows is vital; Russo's research~\cite{russo2024_navigating} shows that LLMs are more readily adopted when they seamlessly fit into current work processes. Trust in the technology is another essential factor, as highlighted by Khojah et al.~\cite{khojah2024beyond}, who found that professional reliance on LLMs for guidance and problem-solving depends heavily on the trustworthiness of these tools.

\steSummaryBox{\faList \hspace{0.05cm} Related Work: Summary and Research Gap.}{The aforementioned studies highlight the transformative potential of LLMs in software development, from enhancing code generation and summarization to improving user interactions and professional task efficiency. They also emphasize the importance of addressing educational and ethical considerations to ensure equitable and effective adoption. Building on this prior research, our study extends the existing knowledge by (1) identifying the most critical factors from the well-established Unified Theory of Acceptance and Use of Technology (UTAUT2) model that influence the adoption of LLMs, and (2) examining the moderating role of cultural values in this adoption process, thereby providing concrete insights applicable to similar research domains.}

\section{Hypothesis and Theory Development}
\label{sec_theory}

This study aims to address two related research gaps. The first gap involves identifying the main individual factors influencing the adoption of LLMs in software development and engineering, a topic not yet thoroughly explored. The second gap focuses on understanding the role of individual cultural values in the adoption process of LLMs.

To achieve our objectives, we developed a set of hypotheses grounded in two theoretical frameworks: the Unified Theory of Acceptance and Use of Technology (UTAUT2)~\cite{Venkatesh_2012_UTAUT2}, widely used to explain technology adoption, and Hofstede's Cultural Dimensions framework~\cite{hofstede_1980_cultural_model_1, hofstede_2011_cultural_model_2}, which decomposes the complex concept of culture into distinct constructs. Below, we explain these two models and the hypotheses derived from them, tailored to the context of LLMs adoption in software development.

\subsection{The Unified Theory of Acceptance and Use of Technology}

\subsubsection{The Theoretical Framework}

Several theoretical models have been developed to predict the adoption and use of technology. One such model is the Unified Theory of Acceptance and Use of Technology (UTAUT), created by Venkatesh et al.~\cite{Venkatesh_2003_UTAUT} to forecast technology acceptance within organizational contexts. UTAUT builds upon the key constructs of eight preexisting models that span fields from human behavior to computer science. It identifies four primary factors influencing the intention to use (BI) and actual usage (UB) of information technology: the belief that using the system will improve job performance (Performance Expectancy, PE), the ease of using the system (Effort Expectancy, EE), the perception that organizational and technical infrastructures are in place to support system use (Facilitating Conditions, FC), and the perception that important others believe the system should be used (Social Influence, SI).

Despite UTAUT's broad acceptance, Venkatesh et al. later introduced UTAUT2, an extended version of the previous model with three additional constructs aimed at capturing aspects related to the user in the role of client and customer rather than simple adopters of technology~\cite{Venkatesh_2012_UTAUT2}. The three new constructs in UTAUT2 are the degree of pleasure or enjoyment derived from using a technology (Hedonic Motivation, HM), the user's cognitive trade-off between the perceived benefits of the technology and the monetary cost of using it  (Price Value, PV), and the extent to which people tend to perform behaviors automatically due to learning (Habit, HB).

\subsubsection{Motivation and Adoption Choices}

Since our aim was to investigate the adoption of LLMs in software engineering and the moderating role of culture, UTAUT2 was an obvious choice. First, UTAUT2 is a widely used and validated instrument for examining technology adoption in numerous contexts~\cite{tamilmani_2021_UTAUT2_LR, srite2006_national_culture_TAM,venkatesh2000don, Venkatesh_2003_UTAUT, Venkatesh_2012_UTAUT2}. Compared to its predecessors (TAM~\cite{davis1989_TAM} and UTAUT~\cite{Venkatesh_2003_UTAUT}), it encompasses a broader set of individual-level factors that capture various aspects of technology adoption~\cite{Venkatesh_2012_UTAUT2}. Additionally, the social influence and facilitating conditions constructs enable us to account for environmental factors that may impact the adoption process~\cite{Venkatesh_2012_UTAUT2}. Lastly, UTAUT2 allowed us to rely on well-established and validated measurement instruments, as well as contextualize our results within a large body of literature using the same theoretical framework.

In our study, we employed the UTAUT2 model to examine the developers' adoption of LLMs in software engineering. Our primary objective was to explore the influence of individual cultural values on this adoption. The standard version of UTAUT2 also involves three moderating variables, i.e., age, gender, and experience. For model parsimony reasons, previous works~\cite{srite2006_national_culture_TAM, baptista_2015_culture_and_mobile_banking} investigating cultural values in UTAUT decided not to include these three moderating constructs. Thus, we opted for the same choice.\footnote{For reliability reasons and thoroughness, we conducted the analysis using the moderating variables of the original model (UTAUT2). The results, available in our online appendix~\cite{online_appendix}, revealed that the original three moderating variables were not significant.} 

We imposed a conditional approach to incorporating the Price Value construct based on our data collection process. We identified three different modalities through which practitioners might use LLMs: provided by their company, accessed through a free version, or obtained via a personal subscription. The Price Value construct is relevant only in the case of personal subscriptions. Consequently, we collected data on the modality used by participants and decided to include the Price Value construct in our final model test only if the majority of participants fell into the third category.\footnote{Our preliminary exploration revealed that most practitioners adopt the free version of LLMs.}

\subsubsection{Hyptothesis Development} 

In the following, we present the hypothesis we developed for the constructs of the UTAUT2 model.

Performance expectancy refers to the extent to which an individual believes that using a particular technology will enhance their job performance~\cite{Venkatesh_2003_UTAUT}. This concept indicates that software engineers are more likely to adopt new technologies, such as LLMs, if they perceive these tools as beneficial for their software development tasks~\cite{compeau1995computer}. The positive outcomes associated with LLMs, such as increased coding efficiency, improved accuracy, and enhanced problem-solving capabilities, can significantly impact engineers' willingness to integrate these tools into their workflow~\cite{khojah2024beyond, liu2024empirical, kumar2024code, russo2024_navigating}. Given the potential for LLMs to streamline development processes and provide substantial performance benefits, it is reasonable to propose that performance expectancy will be a key determinant in software engineers' intention to adopt these technologies. Therefore, we hypothesize:

\begin{itemize}[leftmargin=1.5cm]
    \item[H1] \textit{Performance expectancy (PE) positively influences software engineers’ intention to adopt (BI) LLMs for software development.}
\end{itemize}

Effort expectancy is defined as the degree of ease associated with the use of a particular technology~\cite{Venkatesh_2003_UTAUT}. It implies that individuals are more inclined to adopt new technologies if they find them easy to understand and use~\cite{davis1989perceived}. For software engineers, the ease of integrating and utilizing LLMs in their development processes can significantly influence their adoption decision. If engineers perceive that using LLMs requires minimal effort and reduces the complexity of their tasks, they are more likely to embrace these tools. Given the importance of usability and reduced cognitive load in technology acceptance, it is expected that effort expectancy will positively influence software engineers' intention to adopt LLMs for software development. Therefore, we hypothesize:

\begin{itemize}[leftmargin=1.5cm]
    \item[H2] \textit{Effort expectancy (EE) positively influences software engineers’ intention to adopt (BI) LLMs for software development.}
\end{itemize}

Social influence is the degree to which an individual perceives that important others believe they should use a particular technology~\cite{Venkatesh_2003_UTAUT}. This concept suggests that individuals are more likely to adopt a new technology if they perceive that influential peers, superiors, or societal norms endorse its use. For software engineers, the encouragement and approval from colleagues, mentors, or industry leaders can significantly impact their decision to adopt LLMs. If engineers observe that LLMs are widely accepted and valued within their professional community, they are more likely to follow suit~\cite{venkatesh2000don}. Given the strong role of peer and social pressure in technology adoption, it is anticipated that social influence will positively affect software engineers' intention to adopt LLMs for software development. Therefore, we hypothesize:

\begin{itemize}[leftmargin=1.5cm]
    \item[H3] \textit{Social influence (SI) positively influences software engineers’ intention to adopt (BI) LLMs for software development.}
\end{itemize}

Price value refers to engineers' cognitive evaluation of the trade-off between the perceived benefits of using LLMs and the associated monetary costs~\cite{Venkatesh_2012_UTAUT2}. As mentioned before, it makes sense to consider such an aspect only if the participant is paying the LLM on their own. The price value is considered positive when the perceived benefits of using LLMs outweigh the monetary costs. Therefore, we hypothesize:

\begin{itemize}[leftmargin=1.5cm]
    \item[H4] \textit{Price Value (PV) positively influences software engineers’ intention to adopt (BI) LLMs for software development.}
\end{itemize}

Hedonic motivation refers to the enjoyment or pleasure experienced when using LLMs for software engineering and development purposes~\cite{Venkatesh_2012_UTAUT2} and plays a significant role in user technology acceptance~\cite{van2004user}. It is reasonable to think that as the entertainament of the activity increases, so does the practitioner's intention to adopt it. Therefore, we hypothesize:

\begin{itemize}[leftmargin=1.5cm]
    \item[H5] \textit{Hedonic Motivation (HM) positively influences software engineers’ intention to adopt (BI) LLMs for software development.}
\end{itemize}

Facilitating conditions refer to the belief that an organizational and technical infrastructure exists to support the use of a particular technology~\cite{Venkatesh_2003_UTAUT}. For software engineers, having access to necessary tools, training, and technical support can significantly influence their intention to adopt LLMs~\cite{russo2024_navigating}. When engineers feel confident that their organization provides sufficient support for integrating and using LLMs, they are more likely to adopt these tools~\cite{thompson1991personal}. Furthermore, facilitating conditions also impact the actual use of LLMs, as the availability of resources and support ensures effective integration into daily workflows. Therefore, we hypothesize:

\begin{itemize}[leftmargin=1.5cm]
    \item[H6a] \textit{Facilitating conditions (FC) positively influence software engineers’ intention to adopt (BI) LLMs for software development.}
    
    \item[H6b] \textit{Facilitating conditions (FC) positively influences the actual use behavior (UB) of software engineers regarding LLMs for software development.}
\end{itemize}

Habit refers to the extent to which people tend to perform behaviors automatically because of learning~\cite{Venkatesh_2012_UTAUT2}. For software engineers, the familiarity and routine use of certain tools can significantly influence their intention to adopt LLMs. When engineers are accustomed to integrating new technologies into their workflows, they are more likely to intend to adopt LLMs due to their habitual acceptance of innovation~\cite{Venkatesh_2012_UTAUT2, russo2024_navigating}. Moreover, habit also affects the actual use behavior, as engineers who regularly incorporate LLMs in their working routine are more likely to consistently use LLMs~\cite{khojah2024beyond, agossah2023llm, draxler2023gender, russo2024_navigating}. Therefore, we hypothesize:

\begin{itemize}[leftmargin=1.5cm]
    \item[H7a] \textit{Habit (HB) positively influences software engineers’ intention to adopt (BI) LLMs for software development.}
    
    \item[H7b] \textit{Habit (HB) positively influences the actual use behavior (UB) of software engineers regarding LLMs for software development.}
\end{itemize}

In line with models based on psychological theories, which assert that individual behavior is predictable and influenced by personal intention, UTAUT2 posits that behavioral intention significantly impacts technology use~\cite{Venkatesh_2003_UTAUT}. Therefore, we hypothesize:

\begin{itemize}[leftmargin=1.5cm]
    \item[H8] \textit{Behavioral Intention (BI) to use LLMs positively influences the actual use behavior (UB) of software engineers regarding LLMs for software development.}
\end{itemize}

\subsection{The Role of Individual Cultural Values in Technology Adoption}

\subsubsection{Cultural Values and Frameworks}
Geert Hofstede has defined culture as \textit{“The programming of the human mind by which one group of people distinguishes itself from another group.”}~\cite{hofstede_1980_cultural_model_1}. The cultural background of individuals influences most of the choices made by humans in a plethora of aspects of everyday life. Undoubtedly, culture represents a complex topic to investigate and formalize, raising an (apparently) unsurmountable wall for researchers eager to embark on this path. To overcome this challenge, researchers from the cross-cultural research field started to conduct empirical investigations to identify values (or dimensions) capable of characterizing individuals' cultural backgrounds and differentiating them. This research agenda led to the definition of the so-called \textit{cultural dimension frameworks}, sets of dimensions, each representing values and behaviors representing one of the faceted aspects of culture~\cite{hofstede_1980_cultural_model_1, hofstede_2011_cultural_model_2}. Thus, this study adopted Hofstede's cultural dimensions to conceptualize individual-espoused cultural values.

Hofstede's cultural framework is composed of six cultural value dimensions~\cite{hofstede_1980_cultural_model_1, hofstede_2011_cultural_model_2}. In this study, we considered only five of Hofstede's cultural dimensions, excluding the sixth dimension, i.e., \textit{Indulgence vs. Restraint}. The reasons for this exclusion are twofold: (1) there is a lack of studies exploring its role as a factor for hypothesis formulation, and (2) there are no validated scales available to measure it at the individual level, as existing scales primarily assess it at the national level.

\begin{description}[leftmargin=0.3cm]
    \item[Power Distance Index (PDI).] This dimension measures the extent to which the less powerful members of a society accept and expect an unequal distribution of power. Individuals in high power distance cultures demonstrate strong obedience to and dependence on authority figures. They also tend to adhere closely to organizational norms.

    \smallskip
    \item[Individualism vs. Collectivism (COL).] This dimension assesses the degree to which individuals are integrated into groups. High COL indicates people prefer to take care of themselves and their immediate families rather than integrate themselves into strong, cohesive groups that offer protection in exchange for loyalty.

    \smallskip
    \item[Masculinity vs. Femininity (MAS).] This dimension contrasts societal preferences for achievement, heroism, assertiveness, and material success (high MAS) with preferences for cooperation, modesty, caring for the weak, and quality of life (low MAS).

    \smallskip
    \item[Uncertainty Avoidance Index (UAI).] This dimension reflects the degree to which members of a society feel uncomfortable with uncertainty and ambiguity. Individuals with higher levels of uncertainty avoidance exhibit elevated anxiety, perceive uncertainty as a threat, prefer familiar situations, and show intolerance towards unconventional ideas. They tend to be risk-averse and resistant to innovation and change.

    \smallskip
    \item[Long-Term vs. Short-Term Orientation (LTO).] This dimension measures the extent to which a society values long-term commitments and respect for tradition. Individuals with a long-term orientation are known for their forward-thinking mindset, receptiveness to new ideas, and a reduced emphasis on traditional practices.

    \smallskip
    \item[Indulgence vs. Restraint (IVR).] This dimension gauges the degree of freedom allowed by societal norms in fulfilling human desires. Individuals in indulgent societies are characterized by a free expression of their desires and a tendency to enjoy life and have fun, while those in restrained societies are more likely to control and regulate the fulfillment of their needs through strict social norms.
\end{description}

\subsubsection{Hyptothesis Development} 

From the various proposed frameworks, Hofstede's emerged as one of the most adopted in the field of software engineering research~\cite{SE_and_Cultural_aspects, hofstede_use_1, cultural_aspects_teaching}. Hofstede's cultural values have also been used in conjunction with the construct of UTAUT2 to investigate the potential influence of culture on the adoption of technology~\cite{tamilmani_2021_UTAUT2_LR, baptista_2015_culture_and_mobile_banking, lai_2016_culture_and_teaching_both_approaches, nistor_2014_culture_and_learning_tools_computers}. Some papers from the management research field explored the influence of cultural values as \textit{moderators} for the UTAUT2 constructs~\cite{tamilmani_2021_UTAUT2_LR, baptista_2015_culture_and_mobile_banking, lai_2016_culture_and_teaching_both_approaches, nistor_2014_culture_and_learning_tools_computers}; a \textit{moderator} is a construct able to change the direction of correlation relationships~\cite{hair_2014_PLS}. When the moderator amplifies or strengthens the relationship between the independent and dependent variables, it is referred to as positive moderation; conversely, when the moderator weakens or diminishes this relationship, it is referred to as negative moderation. Since this is a well-accreditated way to integrate the two models (UTAUT2 and Hofstede dimensions), we adopted such a choice. In the following, we present the dimensions composing the framework and how we hypothesized they influence the adoption of LLMs for software engineering, supporting our claims using relevant literature.

Previous research has demonstrated that uncertainty avoidance (UAI) positively moderates the effect of social influence and negatively moderates the effect of performance expectancy on technology usage~\cite{straub1997testing}. For individuals with high uncertainty avoidance, given their resistance to change, social influence plays a more significant role in shaping their adoption intentions~\cite{alhirz2015cultural, hwang2012investigating, srite2006role}. Moreover, their inclination to avoid unstructured or uncertain situations can counteract the perceived usefulness of the technology~\cite{hwang2012investigating}. Furthermore, as Hofstede defines it, UAI is strictly related to people's intention to exit their comfort zone. LLMs are still a young technology, and there is still discussion on the data usage used by the companies developing them. Thus, for individuals with high uncertainty avoidance, intention to use a new technology may be less likely to result in actual use. For the reasons mentioned above, we hypothesized that:

\begin{itemize}[leftmargin=1.5cm]
    \item[H9] \textit{The UAI value of software engineers would moderate the influence of (H9a) Performance Expectancy on Behavioral Intention, (H9b) Social Influence on Behavioral Intention, and (H9c) Behavioral Intention on Use Behavior.}
\end{itemize}

Long-term orientation value (LTO) is related to the tendency of people to take into account future and prospective events in decision-making~\cite{hofstede_2011_cultural_model_2}. Previous research demonstrated that LTO could have a significant moderating effect when integrated into the Technology Acceptance Model. Individuals with a strong long-term orientation may downplay the importance of Hedonic Motivation when considering the use of technology, as they prioritize future benefits over immediate pleasure~\cite{zhou2015toward}. Research indicates that the greater an individual's long-term orientation, the more significantly Performance Expectancy affects their intention to use technology~\cite{lee2013innovation, zhou2015toward}. Additionally, those with a long-term focus are more likely to see their behavioral intentions translate into actual actions, as their commitment to long-term goals strengthens their intentions. Therefore, we propose the following hypothesis:

\begin{itemize}[leftmargin=1.5cm]
    \item[H10] \textit{The LTO value of software engineers would moderate the influence of (H10a) Hedonic Motivation on Behavioral Intention, (H10b) Performance Expectancy on Behavioral Intention, and (H10c) Behavioral Intention on Use Behavior.}
\end{itemize}

People's attitudes to being collectivistic or individualistic largely determine the importance of social influence on choices. Collectivistic people naturally tend to be primarily influenced by their peers; conversely, performance expectancy plays a minor role~\cite{abbasi2015impact, ford2003information}. Moreover, collectivists tend to consider more effort expectancy when adopting a new technology because collective resource and skill utilization can enhance effort expectancy~\cite{abbasi2015impact, childers2001hedonic, markus2014culture}. Furthermore, easier technology use aligns with group decisions and supports individuals' sense of belonging ~\cite{abbasi2015impact}. Additionally, Fagih and Jaradat~\cite{faqih2015assessing} reported that a high collectivistic value negatively moderates the relationship between the intention to use and actual technology use. Consequently, we hypothesize that:

\begin{itemize}[leftmargin=1.5cm]
    \item[H11] \textit{The COL value of software engineers would moderate the influence of (H11a) Performance Expectancy on Behavioral Intention, (H11b) Effort Expectancy on Behavioral Intention, (H11c) Social Influence on Behavioral Intention, and (H11d) Behavioral Intention on Use Behavior.}
\end{itemize}

As for Individualism vs. Collectivism, the Power Distance (PDI) attitude is closely related to Social Influence. Indeed, people with high power distance values are more likely to follow what their superiors or authoritative figures command them to do~\cite{lee2015examining, lin2014investigation}. The same discourse can be applied to the intention to use technology; people with high power distance could be pushed in their intention to use an instrument by their peers' suggestions~\cite{webb2006does}. For such resons, we hypothesize that: 

\begin{itemize}[leftmargin=1.5cm]
    \item[H12] \textit{The PDI value of software engineers would moderate the influence of (H12a) Social Influence on Behavioral Intention and (H12b) Behavioral Intention on Use Behavior.}
\end{itemize}

Masculinity vs Femininity is the value reported as the less influential in technology adoption. Some papers demonstrated that people with high MAS values could amplify the effect of performance expectancy on Behavioral Intention~\cite{venkatesh2010unified}; this is reasonable since masculinity is generally associated with competitiveness. Thus, we made the following hypothesis:

\begin{itemize}[leftmargin=1.5cm]
    \item[H13] \textit{The MAS value of software engineers would moderate the influence of (H13a) Performance Expectancy on Behavioral Intention and (H13b) Behavioral Intention on Use Behavior.}
\end{itemize}

\begin{figure}
    \centering
    \includegraphics[width=1\linewidth]{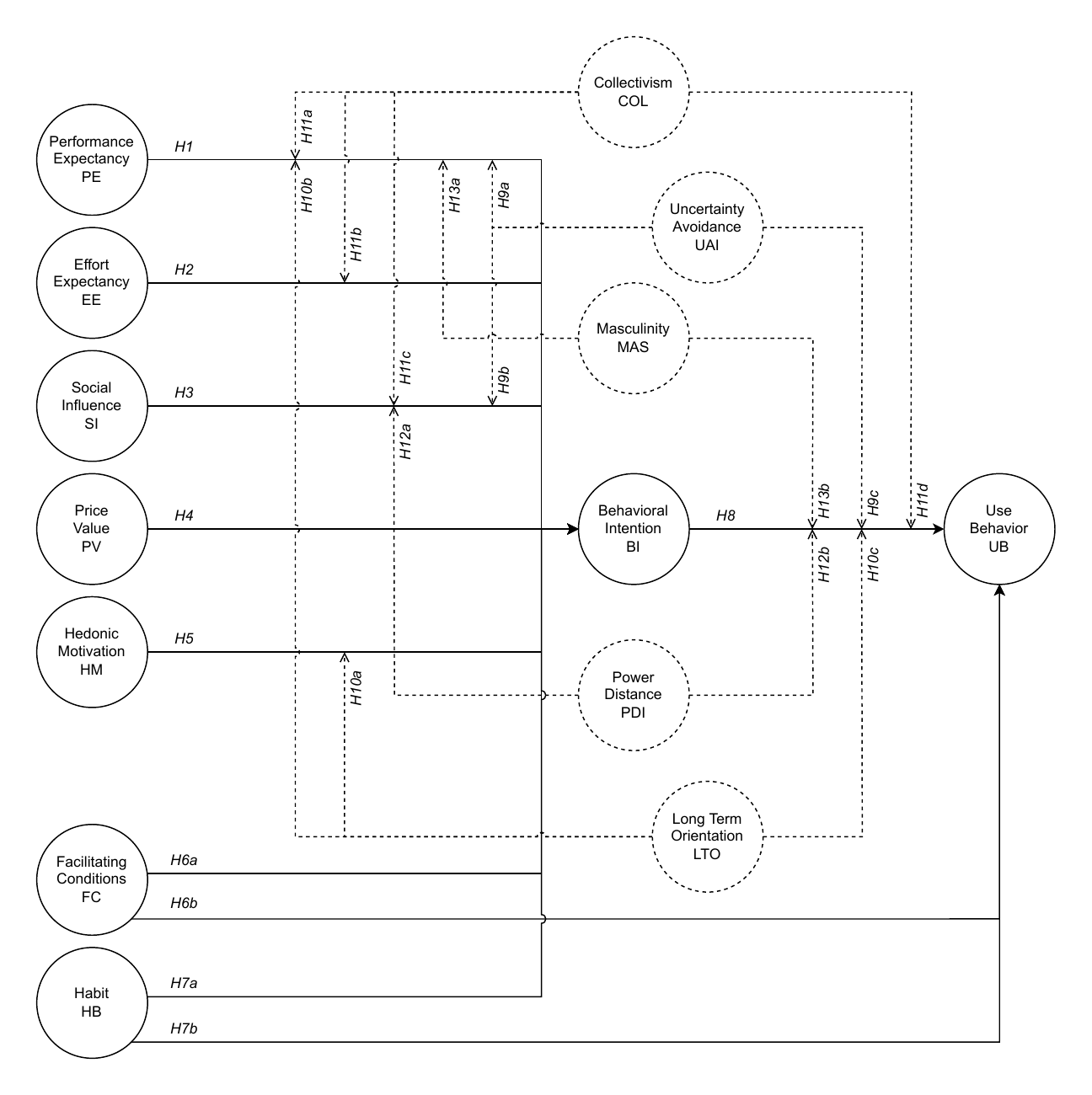}
    \caption{Theoretical Model and Hypothesis.}
    \label{fig_structural_model}
\end{figure}

We integrated the hypothesis mentioned above with the one of UTAUT2 to investigate the adoption of large language models for Software Engineering purposes. Thus, we developed the model shown in Figure \ref{fig_structural_model}. In the next section, we explain the research steps to test our hypothesis, thus our theoretical model, and answer the research question guiding the investigation.

\section{Research Design and Theory Validation}
\label{sec_method}

To test our hypotheses and uncover the role of culture in the adoption of LLMs for software development, we conducted a survey study targeting a sample of software engineers. We collected data using a set of questionnaires, each consisting of items validated by existing literature. Subsequently, we applied \textit{Partial Least Squares-Structural Equation Modeling (PLS-SEM)}~\cite{hair_2014_PLS} to test our hypotheses and validate the theoretical framework described in the previous section~\ref{sec_theory}.

Partial Least Squares-Structural Equation Modeling (PLS-SEM) is a statistical technique used to analyze complex relationships between complex constructs~\cite{hair_2014_PLS}. It combines factor analysis and multiple regression, allowing researchers to model the relationships among multiple independent and dependent constructs simultaneously. PLS-SEM is particularly useful for predictive analysis and theory development when the research model is complex, the sample size is small, or the data distribution is non-normal. It focuses on maximizing the explained variance of the dependent constructs rather than fitting the model to the data.

\begin{figure}
    \centering
    \includegraphics[width=1\linewidth]{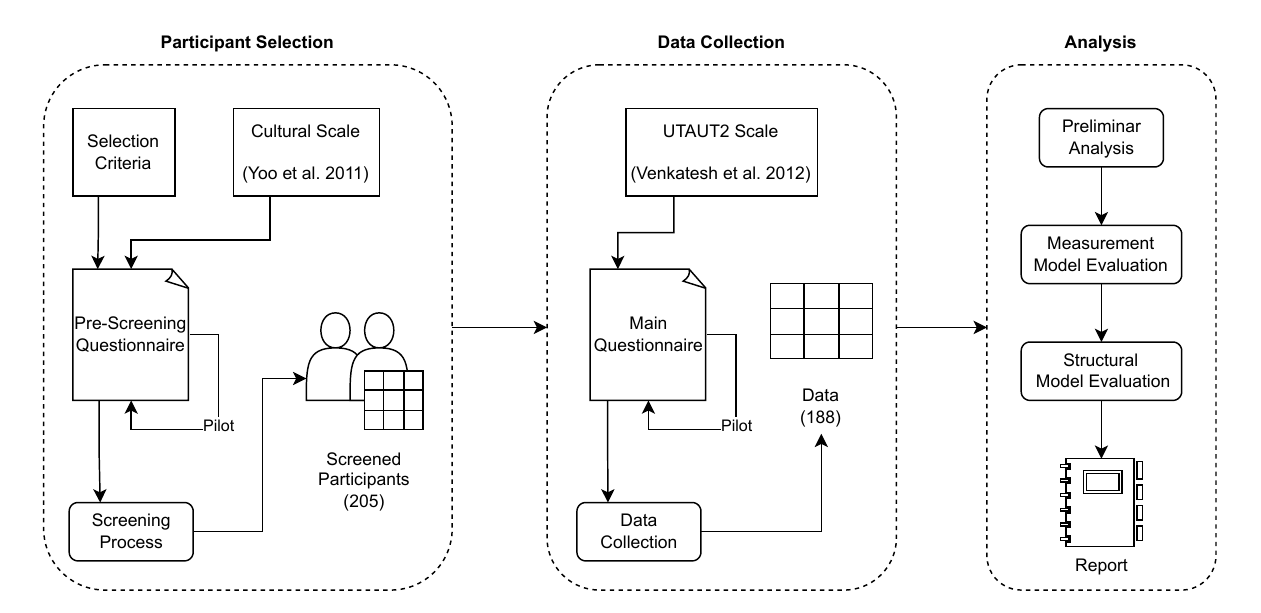}
    \caption{Research Method.}
    \label{fig_research_method}
\end{figure}

Figure \ref{fig_research_method} reports the methodological steps we defined for conducting our investigation. First, we identified a set of criteria for selecting the study participants, as well as the necessary number based on the theoretical model complexity and G*Power~\cite{faul_2009_GPower}. Moreover, starting from the constructs of the theoretical model, we identified validated instruments in the literature to measure them through questionnaires~\cite{Venkatesh_2012_UTAUT2, yoo_2011_CVSCALE_measuring_culture}. As a first step, we developed a questionnaire to screen our participants and collect data on their cultural values. Then, we administered a second questionnaire to the participants who passed the scrutinizing phase to measure the UTAUT2 constructs. Last, we used PLS-SEM to analyze the data and answer our research questions. Regarding the questionnaire development, we used guidelines in the literature and conducted iterative pilots to ensure their quality~\cite{kitchenham2008_PersonalOpinionSurveys, andrews2007_survey_guidelines}. Moreover, since our study involved human participation, we sought and received approval from the ethical board committee of Aalborg University, where this research has been conducted. 

The section is organized as follows. First, in Section \ref{sec_survey_data_collection} we summarize the data collection protocol and the questionnaire employed. Then, in Section \ref{sec_measurement_materials}, we describe the indicators (i.e., the questions) used to measure the constructs in our theoretical model, specifically the UTAUT2~\cite{Venkatesh_2012_UTAUT2} and Cultural Values~\cite{yoo_2011_CVSCALE_measuring_culture} constructs. Next, in Section \ref{sec_participants}, we outline the criteria used to identify the target population for our survey study. This is followed by a description of participants demographics in Section \ref{sec_demographics}. Finally, we provide details on the analysis method in Section \ref{sec_analysis}.

\subsection{Survey and Data collection}
\label{sec_survey_data_collection} 

To collect data, we developed two questionnaires: the first one (from now on referred to as the “Pre-Screening questionnaire”) aimed at identifying the ideal participants for our study from the ones already filtered using the Prolific platform; the second one (from now on referred to as the “Main questionnaire”) aimed to measure the UTAUT2 constructs in the population obtained from the pre-screening survey.

To design both surveys, we relied on the well-known guidelines developed by Kitchenham and Pfleeger~\cite{kitchenham2008_PersonalOpinionSurveys}, along with Andrews et al.~\cite{andrews2007_survey_guidelines}, which are widely recognized in software engineering research. Moreover, we adopted the SIGSOFT Empirical Standard for Questionnaire Surveys to design the investigation~\cite{ralph_2020_empirical_standards}. The questionnaires were made completely anonymous, and we inserted an intro description that provided all the details that were useful to participants in understanding what they had to do. Moreover, we included a closing question asking for any feedback on the questionnaire. Furthermore, for participants’ reliability, we included attention check questions aimed at verifying it.

The questionnaires were designed using Qualtrics, a professional platform for designing questionnaires. We took advantage of various instruments provided by the platform: we checked if the respondent were bots and bias patterns in their responses, and we randomized the order of the questions measuring the constructs to reduce response bias.

We conducted an iterative pilot strategy to (1) test the quality and comprehensibility and (2) estimate the completion time of both surveys. First, we performed three pilots with 10 researchers from our network; after each pilot, we modified the survey to fix typos and improve them according to the received feedback. Then, we conducted a pilot (for each questionnaire) directly in Prolific; we administered the pre-screening questionnaire to 5 participants who met our criteria, and then we did the same for the main questionnaire (we obtained 4 answers since one of the participants did not answered the second survey). This last pilot did not lead to any change in the questionnaire. Thus, we included the answers of the pilot participants in the final dataset.

The pre-screening questionnaire aimed at collecting demographic information, participant reliability, and participants’ programming skills and experience. Moreover, we included in the pre-screening questions for measuring cultural values constructs. It was estimated to have a duration of 6 minutes and was administered on 5 June 2024; it took two days to gather 200 answers.

The main questionnaire contained questions to measure the UTAUT2 constructs. It was estimated to have a duration of 5 minutes and was administered on 7 June 2024; it took 5 days to gather 184 answers.

Since our study involved human participants, we sought and received approval from the Aalborg University Research Ethics Committee in May 2024. All participants were over 18 years old, provided informed consent before participating, and were informed of their right to withdraw at any time. The questionnaire and additional information are available in our online appendix~\cite{online_appendix}.

\subsection{Measurement Instruments}
\label{sec_measurement_materials} 

As previously mentioned, data collection was conducted through a survey study. All constructs in the theoretical model described in Section \ref{sec_theory} were measured at the individual level using items validated in the literature, ensuring the reliability of our measurement process. We explain the items used in the following sections, with a comprehensive table of all items and references provided in the Appendix \ref{AppendixA_label} (Table \ref{table_appendix_ind}).

Questionnaire items measuring the core constructs of the UTAUT2 model were adapted from the original authors~\cite{Venkatesh_2012_UTAUT2}. The dependent variable, \textit{Use Behavior} (UB), was measured using a single-item frequency scale. The eight predictors, including seven used in the final analysis and \textit{Price Value} (excluded due to missing values), were measured using a 7-point Likert scale. These predictors included \textit{Performance Expectancy} (PE, measured with 5 items), \textit{Effort Expectancy} (EE, measured with 6 items), \textit{Social Influence} (SI, measured with 5 items), \textit{Hedonic Motivation} (HM, measured with 3 items), F\textit{acilitating Conditions} (FC, measured with 4 items), \textit{Habit} (HB, measured with 4 items), \textit{Price Value} (PV, measured with 3 items), and \textit{Behavioral Intention} (BI, measured with 3 items).
In addition to these items, as mentioned in Section \ref{sec_theory}, we included a conditional question asking the modality to which the participants to the questionnaire accessed LLMs (free, paid by their company, bought by themselves). We asked for the \textit{Price Value} indexes only in the second case since, in the others, it would make no sense.

Regarding the measurement of cultural values, Hofstede developed a framework that decomposes culture into specific constructs and provides a questionnaire instrument to measure these constructs~\cite{hofstede_2011_cultural_model_2}. However, Hofstede noted that these instruments are validated at the group level, effectively capturing the cultural characteristics of groups rather than individuals. Moreover, the scores provided by Hofstede are associated with country-level data, meaning that using them to characterize individuals based on their country of origin can lead to incorrect results. Therefore, the instrument provided by Hofstede cannot be used to characterize the cultural values of an individual, as was the case in our study.

Given these limitations, various researchers have developed their own scales based on Hofstede's dimensions to measure cultural values at the individual level. In this study, we adopted the scale proposed by Yoo et al.~\cite{yoo_2011_CVSCALE_measuring_culture}, known as CVSALE scale, which has been extensively validated for this purpose. The questionnaire items were measured on a 5-point Likert scale. For the \textit{Long Term Orientation} dimension, a score of 1 indicated “very unimportant,” and a score of 5 indicated “very important.” For all other dimensions, a traditional agreement Likert scale was used. Specifically, we measured \textit{Power Distance} (PDI, measured with 5 items), \textit{Uncertainty Avoidance} (UAI, measured with 5 items), \textit{Collectivism} (COL, measured with 6 items), \textit{Long Term Orientation} (LTO, measured with 6 items), and \textit{Masculinity} (MAS, measured with 4 items). In addition to the variables mentioned above, some demographic background data were collected, including the participants’ age, gender, working role, and years of experience in the software industry. The rationale was that we wanted to characterize our sample with respect to other survey works in the literature.

\subsection{Participants}
\label{sec_participants} 

To collect data for our investigation, we conducted a survey study. We used a cluster sampling strategy through Prolific, an academic data collection platform, to recruit participants.\footnote{Prolific (\url{www.prolific.com}) [May 2024]} Prolific enabled us to specify a set of criteria to filter potential participants according to our requirements. Below, we explain the criteria imposed for participant selection, the rationale behind them, and how we ensured these criteria were met through Prolific or other ways.

Based on our objectives, we defined the ideal participant profile. Since we aimed to gather software engineers’ opinions on using LLMs for software development, we sought practitioners proficient in computer programming, employed full-time or part-time in the software industry, and working as software engineers or in similar roles. Prolific allowed us to set filters for each of these criteria. Additionally, we titled the survey \textit{“[Software Engineering] Pre-Screening — The Role of Culture in the Adoption of LLMs for Supporting Software Engineering”} and clearly stated our goal in the opening description. This helped ensure that only those meeting the pre-screening criteria, even if not actively working as software engineers, participated.

To ensure participants were reliable and capable of answering our questionnaire correctly, we added two further filters. First, participants had to be fluent in English, as the questionnaire was in English. Second, participants needed a 100\% approval rate for previous surveys. In Prolific, a researcher can reject a participant’s response based on various criteria and checks, and the approval rate represents the percentage of a participant’s submissions that have been approved by researchers, indicating their reliability and consistency in providing quality responses.

In addition to the criteria mentioned, we included questions developed by Danilova et al.~\cite{danilova2021_developers_questions} to assess participants' programming knowledge, ensuring competency in programming. Finally, since our focus was on LLMs, we sought participants with at least some familiarity with LLMs in a work context. To assess this, we included a question about the frequency of LLM use.

Based on the criteria mentioned above, Prolific identified a potential pool of 849 participants. The minimum sample size was determined by conducting an a priori power analysis using G*Power~\cite{faul_2009_GPower}. With an effect size of 15\%, a significance level of 5\%, and a power of 95\%, the smallest required sample size for twelve predictors was calculated to be 184 participants. To account for potential attrition between the pre-screening and the main survey, we initially collected 205 responses for the pre-screening survey; for the main one, we gathered 188 responses. The temporal gap between the two surveys allowed us time to screen participants thoroughly. Anticipating that some would not continue to the main survey, we began with a larger number of participants to ensure a sufficient sample size for our analysis.

\subsection{Participants Demographics}
\label{sec_demographics} 

Our survey featured a varied group of 188 participants, including 73.4\% men, 25.5\% women, and 1.1\% non-binary individuals. The respondents work in 13 distinct countries, with the highest numbers of participants from the United Kingdom (27.1\%), the United States (26.6\%), Canada (10.1\%), Germany (9.6\%), and Spain (6.4\%). Other countries represented in the survey include Italy, the Netherlands, France, Austria, Ireland, Australia, Switzerland, and Sweden, reflecting the diverse backgrounds of our respondents. In terms of born country, the majority of respondents were born in North America (28.7\%), with a significant number from the United States (22.3\%) and Canada (6.4\%). Western Europe also had a substantial representation at 28.2\%, primarily from the United Kingdom (19.7\%), Germany (3.7\%), and Spain (3.2\%). Asian countries accounted for 17.0\% of the participants, including India (3.2\%), and China (2.1\%). Eastern Europe contributed 2.7\% to the survey, with notable mentions from Bulgaria and Hungary. Additionally, 4.8\% of participants were born in African nations, 1.1\% in South America, and 1.1\% in Oceania. This wide-ranging geographic distribution underscores the global nature of our respondent pool.

Our survey featured a range of work positions among the 188 respondents. The most common roles were Software Developer / Programmer (33.1\%) and Software Engineer (22.8\%). Other significant positions included Data Analyst / Data Engineer / Data Scientist (10.6\%), Team Leader (7.5\%), Project Manager (6.6\%), and Tester / QA Engineer (5.6\%). UX / UI Designer and those in other specified roles each accounted for 4.4\%. Additionally, Software Architects comprised 3.4\% of respondents, while CIO / CEO / CTO roles represented 1.6\%.

The survey responses reflect a diverse range of experience levels among participants. In terms of working role experience, the majority have between 1-3 years (34.0\%), followed by 4-6 years (28.7\%), and more than 10 years (20.2\%). A smaller portion of respondents have 7-9 years (11.7\%) and less than 1 year (5.3\%) of experience. When considering experience within the software industry, a significant portion of participants have over 10 years (34.0\%) of experience. This is followed by those with 1-3 years (26.1\%), 4-6 years (23.4\%), 7-9 years (11.2\%), and less than 1 year (5.3\%). These statistics highlight a broad spectrum of expertise within the surveyed group, with a notable presence of both early-career and highly experienced professionals.

\subsection{Analysis Overview}
\label{sec_analysis} 

In terms of analysis, we first conducted a \textbf{preliminary examination} of the data collected. Although PLS-SEM does not impose strict constraints, we checked for missing data, suspicious response patterns, outliers, and data distribution to ensure data quality.

Once the data was validated, we imported the dataset into SmartPLS, a software tool designed for conducting PLS-SEM statistical analysis~\cite{SmartPLS4}.
Here is a summary of how analysis with PLS-SEM works. Due to the complexity of the process, we recommend referring to Hair et al.~\cite{hair_2014_PLS} and Russo and Stol~\cite{russo2021_pls_SLR} for a full and comprehensive explanation.
First, we created the measurement model (also known as the outer model), which combines the construct to be measured with its associated indicators. For each construct in the theoretical model (i.e., UTAUT2 and cultural values), we created a latent variable representing a non-observable construct. We then associated the indicators, i.e., the participants' questionnaire responses, with each corresponding construct. Next, we created the structural model (also known as the inner model), which represents the relationships between the constructs based on our theoretical hypotheses. Finally, we incorporated the moderating effect of cultural values into the model. A moderating effect hypothesizes that the value of one construct can change the direction of the relationship between two other constructs. This was achieved by associating each cultural value construct with the corresponding relationships discussed in our hypotheses.

After constructing both the measurement and structural models within the tool and running the PLS-SEM algorithm, we initially evaluated the measurement models. Given that all our indicators had a reflective relationship with the constructs, we assessed the following:
\begin{itemize}
    \item \textbf{Indicator Reliability}: Ensures that each item consistently represents the intended construct.
    \item \textbf{Internal Consistency Reliability}: Assessed using composite reliability to ensure the consistency of the constructs.
    \item \textbf{Convergent Validity}: Verified that the indicators of a construct share a high proportion of variance through the Average Variance Extracted (AVE).
    \item \textbf{Discriminant Validity}: Confirmed that constructs are distinct from one another.
\end{itemize}

Following the measurement model assessment, we proceeded to evaluate the structural model. We assessed:

\begin{itemize}
    \item \textbf{Collinearity}: Ensured that predictor variables in the model are not highly correlated, using Variance Inflation Factor (VIF).
    \item \textbf{Significance of Relationships}: Determined the statistical significance of path coefficients using bootstrapping.
    \item \textbf{Explanatory Power}: Evaluated how well the model explains the variance in the dependent variables using R-squared values.
    \item \textbf{Predictive Power}: Assessed the model's ability to predict outcomes using measures such as Q-squared.
\end{itemize}

Since our structural model included cultural values as moderators, we performed a moderator analysis. In PLS-SEM, conducting a moderator analysis requires introducing an additional construct for each moderating effect, known as the interaction term. This involves multiplying each standardized indicator of the moderator variable with those of the independent variable and creating a relationship between the interaction term and the dependent variable. Both the moderating construct and the interaction term must be used in the model evaluation: the moderating construct is assessed in the measurement model, while the interaction term is evaluated in the structural model.

To ensure replicability and reliability, we included all materials used during the analysis in the replication package accompanying this paper~\cite{online_appendix}.

\section{Analysis of the Results}
\label{sec_results}

The following section presents the results obtained from the PLS-SEM analysis, offering insights into the structural model's path coefficients, the measurement model's reliability and validity, and the overall model fit. Through this analysis, we aim to understand the causal relationships and underlying patterns within the data, providing a comprehensive evaluation of the hypothesized model.

\subsection{Preliminary Examination}
\label{sec_preliminary_Examination} 

Before conducting the principal PLS-SEM investigation, we performed a preliminary data analysis. Notably, we did not encounter any missing data, which can be attributed to the quality of our questionnaire and the reliability of our sample, ensured by the approval rate filter. Furthermore, we checked for suspicious response patterns in the answers and did not find any. It is also worth noting that none of the participants failed the attention check questions.

Regarding outliers, an analysis using boxplots and statistical tools revealed very few. Since our population meets all the requirements and criteria, we chose not to eliminate the outliers as they represent actual data. Additionally, we examined data distribution for skewness and kurtosis. Our analysis indicated nonnormal distribution in the data for two of the indicators for cultural values. As a general guideline, values beyond -2 and +2 indicate nonnormality. Despite normality is not required in PLS-SEM, nonnormal distribution are in general problematic~\cite{hair_2014_PLS}. For such a reason and since only 2 indicators suffered of this problem, we excluded them from the final analysis.

As a final check, we evaluated the price value construct. The majority of participants (51.9\%) use a free version of the software, 33.7\% have their software provided by their company, and only 14.4\% purchased the software independently. Including the price value construct in our analysis was meaningful only for participants who bought the tool independently. Given that most participants did not, we excluded this construct from the final investigation.

\subsection{Measurement Model Evaluation}
\label{sec_measurement_model_evaluation} 

As a first step in the evaluation of the theoretical model, it is paramount to evaluate the reliability of the constructs of the model~\cite{hair_2014_PLS, russo2021_pls_SLR}. Consequently, we analyze the indicator reliability, internal consistency reliability, convergent validity, and discriminant validity. This section presents the obtained results for each of the steps mentioned above.

\begin{table}
    \centering
    \caption{Outer Loadings of the UTAUT2 constructs. (EE = Effort Expectancy, FC = Facilitating Conditions, HB = Habit, HM = Hedonic Motivation, PE = Performance Expectancy, SI = Social Influence, BI = Behavioral Intention, UB = Use Behavior)}
    \rowcolors{1}{gray!15}{white}
    \begin{tabular}{lllllllll}
    \hline
        \rowcolor{headertable}
        \textcolor{white}{\textbf{Indicators}} & \textcolor{white}{\textbf{EE}} & \textcolor{white}{\textbf{FC}} & \textcolor{white}{\textbf{HB}} & \textcolor{white}{\textbf{HM}} & \textcolor{white}{\textbf{PE}} & \textcolor{white}{\textbf{SI}} & \textcolor{white}{\textbf{BI}} & \textcolor{white}{\textbf{UB}} \\ \hline 
        EE\_1 & 0.769 & ~ & ~ & ~ & ~ & ~ & ~ & ~ \\ 
        EE\_2 & 0.814 & ~ & ~ & ~ & ~ & ~ & ~ & ~ \\ 
        EE\_3 & 0.827 & ~ & ~ & ~ & ~ & ~ & ~ & ~ \\ 
        EE\_4 & 0.831 & ~ & ~ & ~ & ~ & ~ & ~ & ~ \\ 
        EE\_5 & 0.714 & ~ & ~ & ~ & ~ & ~ & ~ & ~ \\ 
        EE\_6 & 0.755 & ~ & ~ & ~ & ~ & ~ & ~ & ~ \\ 
        FC\_1 & ~ & 0.848 & ~ & ~ & ~ & ~ & ~ & ~ \\ 
        FC\_2 & ~ & 0.889 & ~ & ~ & ~ & ~ & ~ & ~ \\ 
        FC\_3 & ~ & 0.730 & ~ & ~ & ~ & ~ & ~ & ~ \\ 
        HB\_1 & ~ & ~ & 0.869 & ~ & ~ & ~ & ~ & ~ \\ 
        HB\_2 & ~ & ~ & 0.786 & ~ & ~ & ~ & ~ & ~ \\ 
        HB\_3 & ~ & ~ & 0.734 & ~ & ~ & ~ & ~ & ~ \\ 
        HB\_4 & ~ & ~ & 0.876 & ~ & ~ & ~ & ~ & ~ \\ 
        HM\_1 & ~ & ~ & ~ & 0.933 & ~ & ~ & ~ & ~ \\ 
        HM\_2 & ~ & ~ & ~ & 0.933 & ~ & ~ & ~ & ~ \\ 
        HM\_3 & ~ & ~ & ~ & 0.858 & ~ & ~ & ~ & ~ \\ 
        PE\_1 & ~ & ~ & ~ & ~ & 0.870 & ~ & ~ & ~ \\ 
        PE\_2 & ~ & ~ & ~ & ~ & 0.915 & ~ & ~ & ~ \\ 
        PE\_3 & ~ & ~ & ~ & ~ & 0.877 & ~ & ~ & ~ \\ 
        PE\_4 & ~ & ~ & ~ & ~ & 0.865 & ~ & ~ & ~ \\ 
        PE\_5 & ~ & ~ & ~ & ~ & 0.744 & ~ & ~ & ~ \\ 
        SI\_1 & ~ & ~ & ~ & ~ & ~ & 0.937 & ~ & ~ \\ 
        SI\_2 & ~ & ~ & ~ & ~ & ~ & 0.951 & ~ & ~ \\
        BI\_1 & ~ & ~ & ~ & ~ & ~ & ~ & 0.889 & ~ \\ 
        BI\_2 & ~ & ~ & ~ & ~ & ~ & ~ & 0.905 & ~ \\ 
        BI\_3 & ~ & ~ & ~ & ~ & ~ & ~ & 0.936 & ~ \\
        UB & ~ & ~ & ~ & ~ & ~ & ~ & ~ & 1.000 \\ \hline
    \end{tabular}
    \label{table_outer_loadings_UTAUT2}
\end{table}

\begin{table}
    \centering
    \caption{Outer Loadings of the Cultural constructs as moderators. (COL = Collectivism, LTO = Long Term Orientation, MAS = Masculinity, PDI = Power Distance Index, UAI = Uncertainty Avoidance Index)}
    \rowcolors{1}{gray!15}{white}
    \begin{tabular}{llllll}
    \hline
        \rowcolor{headertable}
        \textcolor{white}{\textbf{Indicators}} & \textcolor{white}{\textbf{COL}} & \textcolor{white}{\textbf{LTO}} & \textcolor{white}{\textbf{MAS}} & \textcolor{white}{\textbf{PDI}} & \textcolor{white}{\textbf{UAI}} \\ \hline
        COL\_2 & 0.634 & ~ & ~ & ~ & ~ \\ 
        COL\_3 & 0.879 & ~ & ~ & ~ & ~ \\ 
        COL\_4 & 0.913 & ~ & ~ & ~ & ~ \\ 
        COL\_6 & 0.723 & ~ & ~ & ~ & ~ \\ 
        LTO\_2 & ~ & 0.927 & ~ & ~ & ~ \\ 
        LTO\_4 & ~ & 0.668 & ~ & ~ & ~ \\ 
        MAS\_1 & ~ & ~ & 0.796 & ~ & ~ \\ 
        MAS\_2 & ~ & ~ & 0.933 & ~ & ~ \\ 
        MAS\_3 & ~ & ~ & 0.797 & ~ & ~ \\ 
        PDI\_1 & ~ & ~ & ~ & 0.900 & ~ \\ 
        PDI\_2 & ~ & ~ & ~ & 0.815 & ~ \\ 
        PDI\_4 & ~ & ~ & ~ & 0.787 & ~ \\ 
        UAI\_2 & ~ & ~ & ~ & ~ & 0.920 \\ 
        UAI\_5 & ~ & ~ & ~ & ~ & 0.811 \\ \hline
    \end{tabular}
    \label{table_outer_loadings_culture}
\end{table}

\subsubsection{Step 1—Indicator Reliability}
\label{sec_indicator_reliability} 

As indicated in the work of Hair et al.~\cite{hair_2014_PLS}, the first step in assessing the measurement model involves checking the reliability of the indicators, specifically the \textit{outer loadings} of the indicators. High outer loadings on a construct indicate that the associated indicators share a significant amount of commonality, which is captured by the construct.

Guided by a practical rule of thumb, indicators with outer loadings exceeding 0.708 are typically retained in the model, while those below 0.40 are usually discarded. For indicators falling between 0.40 and 0.70, the decision to eliminate them should be based on whether their removal enhances internal consistency reliability or convergent validity.

Table \ref{table_outer_loadings_UTAUT2} and Table \ref{table_outer_loadings_culture} report all the outer loadings for the indicators of our model. As observed, some of the indicators, particularly those associated with cultural values, presented outer loadings less than 0.70. This is not surprising, as similar results have been reported in other studies using cultural values. The constructs represent culture as a contraposition between two ways of thinking and behaving. We eliminated all the indicators with outer loadings less than 0.40 and evaluated the remaining indicators based on the aforementioned rules.

\begin{table}
    \centering
    \caption{Internal Consistency Reliability of the Constructs. The black line in the middle of the table separates the UTAUT2 constructs and the Cultural Values constructs.}
    \rowcolors{1}{gray!15}{white}
    \begin{tabular}{l d{2.2} d{2.2} d{2.2} d{2.2}}
    \hline
        \rowcolor{headertable}
        \textcolor{white}{\textbf{Constructs}} & 
        \multicolumn{1}{l}{\textcolor{white}{\textbf{Cronbach's alpha}}} & 
        \multicolumn{1}{c}{\textcolor{white}{{$\mathbf{\rho_A}$}}} & 
        \multicolumn{1}{c}{\textcolor{white}{{$\mathbf{\rho_c}$}}} & 
        \multicolumn{1}{c}{\textcolor{white}{\textbf{AVE}}} 
        \\ \hline
        Performance Expectancy & 0.908 & 0.918 & 0.932 & 0.733 \\ 
        Effort Expectancy & 0.877 & 0.888 & 0.906 & 0.618 \\ 
        Facilitating Conditions & 0.764 & 0.789 & 0.864 & 0.681 \\ 
        Social Influence & 0.878 & 0.887 & 0.942 & 0.891 \\ 
        Habit & 0.836 & 0.860 & 0.890 & 0.670 \\ 
        Hedonic Motivation & 0.894 & 0.905 & 0.934 & 0.826 \\ 
        Behavioral Intention & 0.897 & 0.897 & 0.936 & 0.829 \\
        \hline
        Long Term Orientation & 0.507 & 0.663 & 0.785 & 0.653 \\ 
        Masculinity & 0.830 & 0.987 & 0.881 & 0.713 \\ 
        Power Distance & 0.788 & 0.853 & 0.873 & 0.698 \\ 
        Uncertainty Avoidance & 0.682 & 0.758 & 0.858 & 0.753 \\ 
        Collectivism & 0.813 & 0.942 & 0.871 & 0.633 \\ 
        \hline
    \end{tabular}
    \label{table_internal_consistency_reliability}
\end{table}

\subsubsection{Step 2—Internal Consistency Reliability}
\label{sec_internal_consistency_reliability} 

The second step involved testing internal consistency reliability, ensuring that the indicators consistently and reliably measure the constructs. To assess this, we referred to three measures: \textit{Cronbach's alpha}, \textit{composite reliability} ($\rho_c$), and the \textit{reliability coefficient} ($\rho_A$). 

The results are presented in Table \ref{table_internal_consistency_reliability}. All the values for these measures exceeded the recommended threshold of 0.60~\cite{hair_2014_PLS}, except for Cronbach's alpha for Long Term Orientation. Nevertheless, we retain this construct for three reasons: (1) the lower value can be attributed to the inherent complexity and variability in measuring long-term orientation, and is coherent with similar works using the CVSCALE~\cite{lai_2016_culture_and_teaching_both_approaches}; (2) the scale used for measuring LTO is largely adopted and validated; (3) the other values to measure Internal Consistency Reliability, i.e., the \textit{composite reliability} ($\rho_c$) and the \textit{reliability coefficient} ($\rho_A$), are above the treshold.

\subsubsection{Step 3—Convergent Validity}
\label{sec_convergent_validity} 

Convergent validity is the extent to which a measure correlates positively with alternative measures of the same construct~\cite{hair_2014_PLS}. Since we selected reflective indicators for all the constructs, we expected that the indicators of the same construct converge or share a high proportion of variance. A standard measure used to assess this is the \textit{average variance extracted} (AVE). In general, an AVE value of 0.50 or higher indicates that the construct explains more than half of the variance of its indicators, which is desirable.

Table \ref{table_internal_consistency_reliability} reports the AVE values for each construct in the model. As it is possible to observe, all the constructs vastly exceed the value of 0.50, allowing us to pass this step confidently.

\begin{table}
    \centering
    \caption{Heterotrait–Monotrait ratio of correlations (HTMT) for Discriminant Validity.}
    \begin{tabular}{l|lllllllllllll}
    \hline
        \textbf{} & \textbf{COL} & \textbf{EE} & \textbf{FC} & \textbf{HB} & \textbf{HM} & \textbf{LTO} & \textbf{MAS} & \textbf{PE} & \textbf{PDI} & \textbf{SI} & \textbf{UAI} & \textbf{UB} & \textbf{BI} \\ \hline
        \textbf{COL} & ~ & ~ & ~ & ~ & ~ & ~ & ~ & ~ & ~ & ~ & ~ & ~ & ~ \\ 
        \textbf{EE} & 0.089 & ~ & ~ & ~ & ~ & ~ & ~ & ~ & ~ & ~ & ~ & ~ & ~ \\ 
        \textbf{FC} & 0.092 & 0.803 & ~ & ~ & ~ & ~ & ~ & ~ & ~ & ~ & ~ & ~ & ~ \\ 
        \textbf{HB} & 0.090 & 0.572 & 0.392 & ~ & ~ & ~ & ~ & ~ & ~ & ~ & ~ & ~ & ~ \\ 
        \textbf{HM} & 0.084 & 0.536 & 0.282 & 0.533 & ~ & ~ & ~ & ~ & ~ & ~ & ~ & ~ & ~ \\ 
        \textbf{LTO} & 0.079 & 0.253 & 0.166 & 0.155 & 0.327 & ~ & ~ & ~ & ~ & ~ & ~ & ~ & ~ \\ 
        \textbf{MAS} & 0.141 & 0.079 & 0.047 & 0.132 & 0.069 & 0.164 & ~ & ~ & ~ & ~ & ~ & ~ & ~ \\ 
        \textbf{PE} & 0.097 & 0.791 & 0.544 & 0.753 & 0.607 & 0.201 & 0.083 & ~ & ~ & ~ & ~ & ~ & ~ \\ 
        \textbf{PD} & 0.136 & 0.223 & 0.149 & 0.117 & 0.307 & 0.072 & 0.363 & 0.258 & ~ & ~ & ~ & ~ & ~ \\ 
        \textbf{SI} & 0.069 & 0.424 & 0.360 & 0.564 & 0.362 & 0.147 & 0.054 & 0.579 & 0.136 & ~ & ~ & ~ & ~ \\ 
        \textbf{UA} & 0.344 & 0.117 & 0.124 & 0.063 & 0.052 & 0.557 & 0.097 & 0.049 & 0.148 & 0.072 & ~ & ~ & ~ \\ 
        \textbf{UB} & 0.132 & 0.499 & 0.424 & 0.587 & 0.305 & 0.057 & 0.057 & 0.530 & 0.204 & 0.365 & 0.025 & ~ & ~ \\ 
        \textbf{BI} & 0.070 & 0.671 & 0.453 & 0.730 & 0.591 & 0.189 & 0.083 & 0.817 & 0.234 & 0.439 & 0.058 & 0.554 & ~ \\ \hline
    \end{tabular}
    \label{table_discriminant_validity_HTMT}
\end{table}

\subsubsection{Step 4—Discriminant Validity}
\label{sec_convergent_validity} 

The last test aimed at checking the discriminant validity, i.e., the extent to which a construct is truly distinct from other constructs by empirical standards. To assess the discriminant validity, Henseler et al.~\cite{henseler2015_HTMT} proposed the \textit{heterotrait-monotrait ratio} (HTMT). HTMT is computed for each construct by means of the PLS-SEM algorithm. Typically, a value above 0.90 indicates a lack of discriminant validity, while a value below 0.85 seems warranted. Moreover, it is suggested to rely on a \textit{bootstrapping} procedure to check if the HTMT values are significant different from the threshold values. Bootstrapping is a nonparametric procedure that allows testing the statistical significance of various PLS-SEM results, such as path coefficients, Cronbach’s alpha, and HTMT values.

 Table \ref{table_discriminant_validity_HTMT} reports the values obtained for the HTMT (in the PLS-SEM computation); most of the values fall beneath the threshold of 0.85, and some are slightly above it. We used SmartPLS to run a bootstrapping procedure with \num{10000} subsamples, one-tailed as the test type, and a 0.05 significance level; also in this case, all the values fall beneath the thresholds. This result suggests that every construct in the model represents a distinct phenomenon and allows us to proceed with assessing the structural model.

\steSummaryBox{\faBarChart \hspace{0.05cm} Measurement Model Evaluation: Summary of the results.}{The measurement model evaluation confirmed the reliability and validity of the constructs. Indicator reliability was established, with most outer loadings exceeding the threshold of 0.708. Internal consistency reliability was supported by Cronbach's alpha, composite reliability, and reliability coefficients, most surpassing 0.60. Convergent validity was verified with AVE values above 0.50 for all constructs. Discriminant validity was demonstrated using HTMT values, all of which were below the critical threshold of 0.85, supported by bootstrapping results. Overall, the constructs exhibit strong reliability and validity, ensuring the robustness of the measurement model.}

\subsection{Structural Model Evaluation}
\label{sec_measurement_model_evaluation} 

After assessing the quality of the measurement model, the next step is to test the structural model. While the measurement model assessment focused on the moderating effect to test the influence of cultural values, this phase centers on the interaction term, which encompasses the concrete impact of the moderating effect~\cite{hair_2014_PLS}.

\subsubsection{Step 1—Collinearity Analysis}
\label{sec_collinearity} 

The first step in assessing the structural model involves verifying the collinearity between the exogenous and endogenous variables. Ensuring that our model does not suffer from multicollinearity issues is crucial for accurate path estimation. The structural model's estimation relies on ordinary least squares (OLS) regressions of each endogenous construct on its corresponding predecessor constructs. Therefore, low collinearity is essential to avoid bias in the path estimation.

To assess collinearity, we employed the \textit{Variance Inflation Factor} (VIF), a widely recognized measure for detecting the presence and severity of multicollinearity in multiple regression models. Generally, a VIF value less than 3 indicates no collinearity and is desirable; however, values less than 5 are also considered acceptable.

In our case, most values were below 3 (the greatest among them was 2.47), with only two slightly exceeding this ideal threshold (PE -> BI and EE -> BI). Consequently, we concluded that multicollinearity is not a significant issue in our model.

\begin{table}
    \centering
    \caption{Significance and Relevance of the UTAUT2 constructs. Significant paths are highlighted in bold. (EE = Effort Expectancy, FC = Facilitating Conditions, HB = Habit, HM = Hedonic Motivation, PE = Performance Expectancy, SI = Social Influence, UB = Use Behavior, BI = Behavioral Intention)}
    \rowcolors{1}{gray!15}{white}
    \begin{tabular}{l d{2.2} d{2.2} d{2.2} d{2.2} d{2.2} c}
    \hline
        \rowcolor{headertable}
        \textcolor{white}{\textbf{Hypotheses}} &
        \multicolumn{1}{l}{\textcolor{white}{\textbf{Path Coefficients}}} &
        \multicolumn{1}{l}{\textcolor{white}{\textbf{Bootstrap Mean}}} &
        \multicolumn{1}{l}{\textcolor{white}{\textbf{St. Dev}}} &
        \multicolumn{1}{l}{\textcolor{white}{\textbf{T statistics}}} &
        \multicolumn{1}{l}{\textcolor{white}{\textbf{P values}}} &
        \multicolumn{1}{l}{\textcolor{white}{\textbf{Significance}}}
        \\ \hline
        EE $\rightarrow$ BI & 0.087 & 0.097 & 0.081 & 1.074 & 0.283 & ~ \\ 
        FC $\rightarrow$ UB & 0.173 & 0.174 & 0.062 & 2.799 & 0.005 & **\\ 
        FC $\rightarrow$ BI & 0.008 & 0.017 & 0.067 & 0.116 & 0.908 & ~ \\ 
        HB $\rightarrow$ UB & 0.378 & 0.379 & 0.082 & 4.613 & 0.000 & ***\\ 
        HB $\rightarrow$ BI & 0.274 & 0.260 & 0.083 & 3.292 & 0.001 & ***\\ 
        HM $\rightarrow$ BI & 0.106 & 0.115 & 0.070 & 1.513 & 0.130 & ~ \\ 
        PE $\rightarrow$ BI & 0.463 & 0.441 & 0.091 & 5.060 & 0.000 & ***\\ 
        SI $\rightarrow$ BI & -0.034 & -0.024 & 0.057 & 0.600 & 0.549 & ~ \\ 
        BI $\rightarrow$ UB & 0.176 & 0.171 & 0.091 & 1.940 & 0.050 & *\\ 
        \hline
        \rowcolor{white}
        \multicolumn{7}{l}{
        $ 
        {\scriptstyle ***: p < 0.001 ;\:\:\: **: p < 0.01 ;\:\:\: *: p < 0.05}
        $
    }\\
    \end{tabular}
    \label{table_significance_relevance_UTAUT2}
\end{table}

\begin{table}
    \centering
    \caption{Significance and Relevance of the Cultural Values Moderators. The hypotheses listed below the black line in the middle of the table represent the interaction terms generated for the moderating analysis. (COL = Collectivism, LTO = Long Term Orientation, MAS = Masculinity, PDI = Power Distance Index, UAI = Uncertainty Avoidance Index)}
    \rowcolors{1}{gray!15}{white}
    \begin{tabular}{l d{2.2} d{2.2} d{2.2} d{2.2} d{2.2}}
    \hline
        \rowcolor{headertable}
        \textcolor{white}{\textbf{Hypotheses}} & 
        \multicolumn{1}{l}{\textcolor{white}{\textbf{Path Coefficients}}} &
        \multicolumn{1}{l}{\textcolor{white}{\textbf{Bootstrap Mean}}} &
        \multicolumn{1}{l}{\textcolor{white}{\textbf{St. Dev}}} &
        \multicolumn{1}{l}{\textcolor{white}{\textbf{T statistics}}} &
        \multicolumn{1}{l}{\textcolor{white}{\textbf{P values}}} 
        \\ 
        \hline
        COL $\rightarrow$ UB & -0.098 & -0.090 & 0.070 & 1.399 & 0.162 \\ 
        COL $\rightarrow$ BI & -0.031 & -0.043 & 0.063 & 0.491 & 0.624 \\ 
        LTO $\rightarrow$ UB & -0.010 & -0.001 & 0.079 & 0.130 & 0.897 \\ 
        LTO $\rightarrow$ BI & 0.025 & 0.017 & 0.065 & 0.381 & 0.703 \\ 
        MAS $\rightarrow$ UB & -0.044 & -0.043 & 0.079 & 0.555 & 0.579 \\ 
        MAS $\rightarrow$ BI & -0.039 & -0.034 & 0.057 & 0.688 & 0.491 \\ 
        PDI $\rightarrow$ UB & -0.115 & -0.122 & 0.070 & 1.639 & 0.101 \\ 
        PDI $\rightarrow$ BI & -0.061 & -0.072 & 0.057 & 1.056 & 0.291 \\ 
        UAI $\rightarrow$ UB & 0.002 & -0.001 & 0.071 & 0.024 & 0.981 \\ 
        UAI $\rightarrow$ BI & -0.007 & -0.003 & 0.058 & 0.113 & 0.910 \\ 
        \hline
        UAI $\times$ SI $\rightarrow$ BI & -0.067 & -0.060 & 0.060 & 1.112 & 0.266 \\ 
        MAS $\times$ PE $\rightarrow$ BI & 0.021 & 0.021 & 0.052 & 0.393 & 0.694 \\ 
        LTO $\times$ BI $\rightarrow$ UB & -0.035 & -0.030 & 0.068 & 0.515 & 0.607 \\ 
        UAI $\times$ BI $\rightarrow$ UB & -0.021 & -0.018 & 0.065 & 0.328 & 0.743 \\ 
        UAI $\times$ PE $\rightarrow$ BI & -0.038 & -0.046 & 0.060 & 0.642 & 0.521 \\ 
        COL $\times$ SI $\rightarrow$ BI & 0.072 & 0.073 & 0.068 & 1.058 & 0.290 \\ 
        COL $\times$ PE $\rightarrow$ BI & 0.058 & 0.050 & 0.087 & 0.669 & 0.503 \\ 
        PDI $\times$ BI $\rightarrow$ UB & 0.022 & 0.025 & 0.064 & 0.344 & 0.731 \\ 
        LTO $\times$ HM $\rightarrow$ BI & -0.009 & -0.012 & 0.079 & 0.117 & 0.907 \\ 
        LTO $\times$ PE $\rightarrow$ BI & -0.067 & -0.056 & 0.080 & 0.834 & 0.404 \\ 
        MAS $\times$ BI $\rightarrow$ UB & 0.062 & 0.022 & 0.082 & 0.748 & 0.454 \\ 
        COL $\times$ BI $\rightarrow$ UB & -0.064 & -0.079 & 0.074 & 0.861 & 0.389 \\ 
        COL $\times$ EE $\rightarrow$ BI & -0.077 & -0.057 & 0.072 & 1.067 & 0.286 \\ 
        PDI $\times$ SI $\rightarrow$ BI & -0.009 & 0.000 & 0.061 & 0.141 & 0.888 \\ \hline
    \end{tabular}
    \label{table_significance_relevance_culture}
\end{table}

\subsubsection{Step 2—Significance and Relevance of the Relationships}
\label{sec_significance} 

The second step in the analysis involved testing for the significance of the relationships of the structural model as well as for their relevance. Such a test occurs by means of \textit{bootstrapping} (for the significance) and observing the \textit{path coefficients} of the relationships (for the relevance); such path coefficients are standardized (their values can range from -1 to +1) and thus can be analyzed relative to each other; a higher path coefficient indicates a greater relevance.

Regarding the significance, we used a two-tailed bootstrapping method with \num{10000} sub-samples. T-values, p-values, or bootstrap confidence intervals can be used to test the significance. Hair et al.~\cite{hair_2014_PLS} suggest using the last one as a measure. Table \ref{table_significance_relevance_UTAUT2} and Table \ref{table_significance_relevance_culture} report the values for testing the significance of our relationships; for the moderating variables (Table \ref{table_significance_relevance_culture}), one has to see the values associated with the interaction term (the ones expressed as products of two constructs) rather than the one associated with the moderating construct. As it is possible to observe, performance expectancy and Habit have a significant relationship with Behavioral Intention. Moreover, Facilitating Condition, Habit, and Behavioral Intention are significantly related to Use Behavior. Furthermore, none of the cultural values constructs is significant.

We then turned our attention to the relevance of the significant relationships. We did this by closely examining the path coefficients of these paths in the PLS-SEM algorithm computation; such values are also reported in second column of Table \ref{table_significance_relevance_UTAUT2} and Table \ref{table_significance_relevance_culture}. Performance Expectancy is the most relevant construct for the intention to use LLMs in software development, followed by Habit. Moreover, focusing on the concrete use of LLMs, Habit has the greatest relevance, followed by the intention to use the LLM and Facilitating Condition.

\subsubsection{Step 3—Explanatory Power}
\label{sec_explanatory_power} 

The third step of our analysis involved assessing the model's explanatory power, specifically its ability to fit the data by quantifying the strength of the associations~\cite{hair_2014_PLS, russo2021_pls_SLR}. This is generally measured using the \textit{coefficient of determination} (R\textsuperscript{2}) of the dependent variables, which ranges from 0 to 1. Higher R\textsuperscript{2} values indicate greater explanatory power. While there are no universally agreed-upon standards for R\textsuperscript{2} values, in some contexts, a value of 0.10 may be satisfactory, though a commonly accepted benchmark is 0.19 or greater~\cite{chin1998_R_Square_value, raithel2012_explanatory_power_PLS_SEM, hair_2014_PLS}.

Our model yields an R\textsuperscript{2} value of 0.64 for Behavioral Intention and 0.41 for Use Behavior. This indicates that our model explains 64\% of the variance in the intention to use LLMs and 41\% of the actual use. Additionally, since our R\textsuperscript{2} values are below 0.90, we can dismiss concerns about overfitting.

After assessing the coefficient of determination, it is useful to quantify the strength of the relationships using the F\textsuperscript{2} \textit{effect size}. This measure indicates the impact on R\textsuperscript{2} if a specific construct is omitted from the model, thus providing insight into the concrete impact of each construct on the dependent variables.

Focusing on the intention to use LLMs, Performance Expectancy exhibits the greatest effect size (0.17), followed by Habit (0.09). For the actual use of LLMs, Habit has the highest effect size (0.12), followed by Facilitating Conditions (0.04) and Intention Behavior (0.02). These results highlight the predominant role of performance expectancy in influencing the intention to use LLMs and the importance of habitual usage in predicting actual use.

\begin{table}
    \centering
    \caption{Prediction Summary (PLS-SEM) and Comparison with Linear Regression (LM). (UB = Use Behavior, BI = Behavioral Intention)}
    \rowcolors{1}{gray!15}{white}
    \begin{tabular}{l d{2.2} d{2.2} d{2.2} d{2.2} d{2.2}}
    \hline
        \rowcolor{headertable}
        \textcolor{white}{\textbf{Indicators}} &
        \multicolumn{1}{l}{\textcolor{white}{\textbf{Q²\textsubscript{predict}}}} &
        \multicolumn{1}{l}{\textcolor{white}{\textbf{PLS-SEM\_RMSE}}} &
        \multicolumn{1}{l}{\textcolor{white}{\textbf{PLS-SEM\_MAE}}} &
        \multicolumn{1}{l}{\textcolor{white}{\textbf{LM\_RMSE}}} &
        \multicolumn{1}{l}{\textcolor{white}{\textbf{LM\_MAE}}} 
        \\ 
        \hline
        UB & 0.312 & 1.088 & 0.820 & 1.222 & 0.929 \\ 
        BI\_1 & 0.431 & 0.694 & 0.531 & 0.728 & 0.539 \\ 
        BI\_2 & 0.409 & 1.022 & 0.787 & 1.085 & 0.838 \\ 
        BI\_3 & 0.447 & 0.873 & 0.657 & 0.956 & 0.710 \\ \hline
    \end{tabular}
    \label{table_prediction_summary}
\end{table}

\subsubsection{Step 4—Predictive Power}
\label{sec_predictive_power} 

To assess the model's utility for managerial decision-making, we tested whether the results of our PLS-SEM algorithm are applicable beyond the data used in the estimation process. We employed the PLS\textsubscript{predict} procedure~\cite{shmueli2016_PLS_SEM_Predict}, which involves separating the overall dataset into training and holdout samples. The measures used in this analysis include Stone-Geisser's \textit{Q\textsuperscript{2} statistic}, the \textit{mean absolute error} (MAE), and the \textit{root mean square error} (RMSE). These statistics were tested against a benchmark, with Shmueli et al.~\cite{shmueli2016_PLS_SEM_Predict, shmueli2019_PLS_SEM_Predict} proposing the use of linear regression model (LM) results as the benchmark.

Positive Q\textsuperscript{2} values indicate that the model's prediction error is smaller than that of the benchmark. Additionally, lower values of MAE and RMSE suggest that the model possesses high predictive power.

Table \ref{table_prediction_summary} reports the values for the dependent variables in the model. All variables exhibit better values compared to the benchmark, indicating that the PLS-SEM model demonstrates high predictive power.

\steSummaryBox{\faBarChart \hspace{0.05cm} Structural Model Evaluation: Summary of the results.}{The structural model evaluation confirmed the robustness of the proposed model. Collinearity analysis showed acceptable VIF values, indicating no significant multicollinearity issues. Path coefficients and bootstrapping results showed significant relationships, particularly for performance expectancy and habit with behavioral intention, and facilitating conditions, habit, and behavioral intention with use behavior. Furthermore, cultural values moderating effect does not seem to be significant. The model exhibited strong explanatory power with R² values of 0.64 for behavioral intention and 0.41 for use behavior. Predictive power assessment through PLS\textsubscript{predict} indicated high predictive accuracy, with all variables showing better values compared to the linear regression benchmark.}

\section{Discussion and Implications}
\label{sec_discussion}

This study set out to explore the impact of cultural values on the adoption of Large Language Models in software engineering, employing the Unified Theory of Acceptance and Use of Technology framework. Our primary objective was to understand whether cultural background plays a significant role in influencing software engineers' decisions to adopt LLMs, a pertinent inquiry given the increasing integration of generative AI in the field.

In the following section we discussed the statistical results of our investigation and provide a set of implications and lessons learned both for practitioners and researchers.

\begin{table}
    \centering
    \caption{UTAUT2—Summary of Findings and Implications.}
    \rowcolors{1}{gray!15}{white}
    \begin{tabular}{p{0.25\linewidth} p{0.325\linewidth} p{0.325\linewidth}}
    \hline
        \rowcolor{headertable} 
        \textcolor{white}{\textbf{Hypotheses}} & \textcolor{white}{\textbf{Findings}} & \textcolor{white}{\textbf{Implications}} \\ 
        \hline
        H1: Performance Expectancy $\rightarrow$ Behavioral Intention &  \textbf{Supported}. This relationship is the strongest of the model, with a path coefficient of 0.46 and an effect size of 0.17 & Software engineers' intentions to adopt LLMs are heavily influenced by their expectations of the technology's performance.
        \\ 
        
        H2: Effort Expectancy $\rightarrow$ Behavioral Intention & Not supported. The relationship is not significant. & The perceived effort in learning how to use and use LLMs alone does not instigate the intention to adopt the tool in working context.
        \\ 
        
        H3: Social Influence $\rightarrow$ Behavioral Intention & Not supported. The relationship is not significant. & The peers opinion on the use of LLMs in the working context alone does not instigate the intention to adopt the tool.
        \\ 
        
        H4: Price Value $\rightarrow$ Behavioral Intention & Not investigated. The lack of a considerable number of individuals having purchased LLMs with their own money did not allow the hypothesis to be investigated. & N/A
        \\ 
        
        H5: Hedonic Motivation $\rightarrow$ Behavioral Intention & Not supported. The relationship is not significant. & Developers do not intend to adopt LLMs on the basis of the fun and joy their use causes.
        \\ 
        
        H6a: Facilitating Conditions $\rightarrow$ Behavioral Intention & Not supported. The relationship is not significant. & Organizational support and supporting resources do not significantly influence developers’ intention to adopt LLMs tools in the working context.
        \\
        
        H6b: Facilitating Conditions $\rightarrow$ Use Behavior & \textbf{Supported}. This relationship is the third most significant regarding Use Behavior, with a path coefficient of 0.17 and an effect size of 0.04. A mediation analysis revealed that the effect is direct and not mediated by the relationship with Behavioral Intention. & While engineers may not initially consider the availability of resources and support systems in their decision to adopt LLMs, these factors become crucial once they start using the technology.
        \\
        
        H7a: Habit $\rightarrow$ Behavioral Intention & \textbf{Supported}. This relationship is the second most significant regarding Behavioral Intention, with a path coefficient of 0.27 and an effect size of 0.09. & Developers that regularly utilized LLMs in their work strengthen their intention to rely on these tools for their tasks.\\
        
        H7b: Habit $\rightarrow$ Use Behavior & \textbf{Supported}. This relationship is the most significant regarding Use Behavior, with a path coefficient of 0.38 and an effect size of 0.12. A mediation analysis revealed that the effect is direct and not mediated by the relationship with Behavioral Intention. & The habitual use of LLMs can lead to higher adoption rate.\\

        H8: Behavioral Intention $\rightarrow$ Use Behavior & \textbf{Supported}. This relationship is the second most significant regarding Use Behavior, with a path coefficient of 0.18 and an effect size of 0.02. & As expected, the intention to adopt LLMs results in an actual adoption of the technology.\\
        \hline
    \end{tabular}
    \label{table_discussion_UTAUT2}
\end{table}

\subsection{Discussion on Hypotheses Significance}

The findings revealed that cultural values did not have a significant impact on the intention to adopt or the actual adoption of LLMs. In contrast, several constructs from the UTAUT2 model—namely habit, performance expectancy, facilitating conditions, and Behavioral Intention—were found to significantly influence the depdendent variables. These results suggest that individual and organizational factors, rather than cultural backgrounds, play a more critical role in the adoption process of LLMs in software engineering.

With these findings in mind, the following sections will provide a detailed discussion of each UTAUT2 construct and the cultural values construct, examining their relevance and the rationale behind our results.

\subsubsection{On the Direct Effect of Technology Acceptance Constructs}

First, we focused on the results regarding the UTAUT2 constructs. Table \ref{table_discussion_UTAUT2} summarizes our findings.

\textit{Performance Expectancy} emerged as the most influential factor affecting the \textit{behavioral intention to use LLMs} for software engineering. This result underscores the critical role that perceived benefits play in adopting new technologies~\cite{Venkatesh_2003_UTAUT, compeau1995computer}. In the context of software engineering, practitioners are highly motivated by tools that can demonstrably enhance their productivity and efficiency. In line with the related literature~\cite{khojah2024beyond, liu2024empirical, kumar2024code}, with their capability to generate code, assist in debugging, and provide decision support, LLMs directly address core aspects of software engineers' tasks, fulfilling their performance expectations. This alignment with crucial performance goals makes LLMs particularly attractive to users who are constantly seeking ways to improve their work outcomes. Moreover, the clear and immediate benefits of LLMs in streamlining complex processes probably reinforce the positive perception of their utility, leading to a strong intention to integrate these models into daily workflows~\cite{agossah2023llm, khojah2024beyond, russo2024_navigating}. This finding aligns with the broader technology adoption literature, which consistently identifies performance expectancy as a primary driver of behavioral intention, particularly when the technology in question offers significant enhancements to job performance~\cite{nistor_2014_culture_and_learning_tools_computers}. Moreover, this result also aligns with the findings from the work of Russo, who found the capacity of an LLM to integrate into the daily workflow of developers as a principal factor in the intention to use it~\cite{russo2024_navigating}.

\textit{Habit} was identified as the second most influential factor on \textit{behavioral intention} to use LLMs for software engineering, and it also showed outstanding significance. Furthermore, Habit exhibited the highest path coefficient with \textit{actual use behavior}, and mediation analysis revealed that this influence is direct and not mediated by \textit{behavioral intention}. This result highlights the pivotal role of habitual behavior in the adoption and continued use of LLMs and are consistent with previous literature on the matter~\cite{khojah2024beyond, agossah2023llm, draxler2023gender, russo2024_navigating}. The direct influence of Habit on use behavior suggests that when engineers incorporate LLMs into their regular workflows, the technology becomes an integral part of their daily activities. This habitual use is likely driven by the consistent and tangible benefits that LLMs provide, such as increased efficiency and ease of performing repetitive tasks. The strong habitual behavior observed underscores the importance of initial exposure and ease of integration into existing processes, as these factors can lead to the sustained use of LLMs.

\textit{Facilitating Conditions} emerged as the third most significant variable influencing \textit{actual use behavior of LLMs}, with a very strong path coefficient and direct relationship, unmediated by \textit{behavioral intention}. This result underscores the critical role that the availability of resources, support, and infrastructure plays in the practical adoption and utilization of technology~\cite{Venkatesh_2003_UTAUT, thompson1991personal}. Moreover, these findings are consistent with the work of Russo~\cite{russo2024_navigating}, which shows that the facilitating of the integration of LLMs in the development lifecycle is pivotal. Furthermore, this direct relationship indicates that even if engineers do not initially intend to use LLMs, the ease of access and the supportive environment can drive actual use. The lack of a significant relationship between Facilitating Conditions and Behavioral Intention suggests that practical considerations and immediate usability factors outweigh initial user intentions when it comes to the adoption of LLMs~\cite{lai_2016_culture_and_teaching_both_approaches}. In summary, engineers are more likely to use LLMs when they feel adequately supported and equipped to do so, regardless of their prior intention. By ensuring that these conditions are met, organizations can effectively promote the use of LLMs, even among those who may not have initially planned to adopt the technology.

\textit{Behavioral Intention} is the second most significant variable influencing \textit{actual use behavior of LLMs}. This result highlights that a strong intention to use LLMs translates directly into their actual adoption. This aligns with the broader technology acceptance literature, which shows that behavioral intention is a key predictor of actual use~\cite{Venkatesh_2003_UTAUT, Venkatesh_2012_UTAUT2}. By fostering positive intentions through training, awareness programs, and demonstrating the benefits of LLMs, organizations can effectively encourage their adoption and integration into daily workflows. Interestingly, other variables such as \textit{Facilitating Conditions} and \textit{Habit} show even stronger direct effects on actual use than Behavioral Intention does. This suggests that while having the intention to use LLMs is important, the practical aspects of support and resources (Facilitating Conditions) and the integration of LLMs into daily routines (Habit) play an even more critical role in ensuring actual adoption~\cite{russo2024_navigating, khojah2024beyond, agossah2023llm, draxler2023gender}. These findings emphasize the need for organizations to not only build positive intentions but also provide robust support systems and encourage habitual use to maximize the adoption and effective utilization of LLMs.

\textit{Effort Expectancy} was found to be non-significant in its relationship with \textit{behavioral intention to use LLMs in software engineering} despite the broader technology acceptance literature~\cite{Venkatesh_2003_UTAUT, Venkatesh_2012_UTAUT2}. This result suggests that the perceived ease of use of LLMs does not strongly influence engineers' intentions to adopt these tools and finds support in the results of Russo~\cite{russo2024_navigating}. One possible justification for this finding is that software engineers, as highly specialized professionals, might not view the effort required to learn and use LLMs as a major barrier. Given their technical expertise and familiarity with complex tools, they may prioritize the functionality and performance benefits of LLMs over the effort required to use them. Additionally, the strong influence of other factors such as \textit{Performance Expectancy} and \textit{Habit} indicates that the potential improvements in job performance and the integration of LLMs into daily routines are more critical in shaping their adoption intentions; this is also supported by related works investigating LLMs in software development~\cite{russo2024_navigating, khojah2024beyond, agossah2023llm, draxler2023gender}. This highlights the importance of focusing on the tangible benefits and seamless integration of LLMs rather than solely on their ease of use when promoting these tools to software engineers.

\textit{Social Influence} was found to be non-significant in its relationship with \textit{behavioral intention to use LLMs in software engineering}. This result indicates that the opinions and behaviors of colleagues and peers do not strongly impact an engineer's intention to adopt LLMs and is consistent with the work of Russo~\cite{russo2024_navigating}. One possible justification is that software engineers tend to base technology adoption decisions more on their individual assessment of a tool's utility and performance rather than on social factors. This is further explained by the Gartner Hype Cycle's latest report~\cite{Gartner2023_AI}, which indicates that generative AI is still in its nascent stage and perceived as a disruptive technology. Consequently, individual perception plays a more significant role in the intention to adopt such technology. Based on this findings, efforts to promote LLM adoption should focus on demonstrating tangible benefits and improving individual user experiences rather than relying on social influence.

Similarly to the previous factor, Social Influence, \textit{Hedonic Motivation} was found to be non-significant in its relationship with \textit{Behavioral Intention} for LLMs in software engineering. This result suggests that the pleasure or enjoyment derived from using LLMs does not strongly influence engineers' intentions to adopt these tools. This aligns with related research noting that habit and performance perception matter most~\cite{khojah2024beyond, agossah2023llm, draxler2023gender}. One possible explanation is that software engineers prioritize practical and performance-related aspects of tools over the enjoyment factor. Given the professional context and focus on efficiency and productivity, engineers may value tools that enhance job performance and streamline workflows rather than those that provide hedonic satisfaction. Therefore, promoting LLM adoption in this field should emphasize practical advantages and performance enhancements rather than enjoyment.

\begin{table}
    \centering
    \caption{Cultural Values—Summary of Findings and Implications.}
    \rowcolors{1}{gray!15}{white}
    \begin{tabular}{p{0.25\linewidth} p{0.2\linewidth} p{0.45\linewidth}}
    \hline
        \rowcolor{headertable} 
        \textcolor{white}{\textbf{Hypotheses}} & \textcolor{white}{\textbf{Findings}} & \textcolor{white}{\textbf{Implications}} \\ 
        \hline
        H9: Uncertainty Avoidance moderates PE $\rightarrow$ BI, SI $\rightarrow$ BI, and BI $\rightarrow$ UB & Not supported. The relationships are not significant. & The tendency of developers to prefer or avoid uncertainty does not moderate the intention to adopt or the actual use of LLMs in software engineering, suggesting that the disruptive nature of LLMs and their potential to significantly enhance perceived productivity diminishes the impact of individual uncertainty preferences on their adoption.\\ 

        H10: Long Term Orientation moderates HM $\rightarrow$ BI, PE $\rightarrow$ BI, and BI $\rightarrow$ UB & Not supported. The relationships are not significant. & The long-term orientation of developers does not act as a moderator for the intention to adopt or the actual use of LLMs in software engineering, suggesting that the immediate and practical benefits of LLMs, such as enhanced performance and usability, take precedence over long-term cultural values in decision-making. \\

        H11: Collectivism moderates PE $\rightarrow$ BI, EE $\rightarrow$ BI, SI $\rightarrow$ BI, and BI $\rightarrow$ UB & Not supported. The relationships are not significant. & Collectivistic or individualistic tendencies do not moderate the intention to adopt or the actual use of LLMs in software engineering, suggesting that in this field, the practical benefits and enhancements to individual performance and efficiency provided by LLMs overshadow the influence of social and cultural contexts on adoption decisions.\\

        H12: Power Distance moderates SI $\rightarrow$ BI and BI $\rightarrow$ UB & Not supported. The relationships are not significant. & A tendency to rigidly follow hierarchical structures and orders does not moderate the intention to adopt or the actual use of LLMs in software engineering, suggesting that the disruptive nature of LLMs, which significantly enhances perceived productivity, diminishes the impact of hierarchical dynamics on adoption. Moreover, the early stage of generative AI technology may explain the weak influence of hierarchical structures on its adoption.\\

        H13: Masculinity moderates PE $\rightarrow$ BI and BI $\rightarrow$ UB & Not supported. The relationships are not significant. & An orientation toward a competitive or collaborative nature does not moderate the intention to adopt or the actual use of LLMs in software engineering, suggesting that practical considerations and the inherent advantages of LLMs, such as substantial improvements in productivity, are the primary drivers of adoption rather than individual or cultural preferences for competition or collaboration.\\
        \hline
    \end{tabular}
    \label{table_discussion_Culture}
\end{table}

\subsubsection{On the Moderating Effect of Cultural Values}

In this section, we focus on the significance of the cultural values constructs treated as moderators in the UTAUT2 model. Table \ref{table_discussion_Culture} summarizes our findings.

\textit{Power Distance} was hypothesized to moderate the relationship between Social Influence and Behavioral Intention, as well as the relationship between Behavioral Intention and Use Behavior. However, this moderating effect was not significant. One primary reason is that the impact of Social Influence on the intention to adopt Large Language Models (LLMs) is negligible, thus rendering any potential moderating effect insignificant. Additionally, the disruptive nature of LLMs, which significantly enhances perceived productivity~\cite{agossah2023llm, khojah2024beyond}, diminishes the impact of hierarchical dynamics on adoption.
Moreover, the nascent stage of generative AI technology places it in the early phase of the Gartner Hype Cycle, as indicated in the 2023 Gartner report on the state of Artificial Intelligence~\cite{Gartner2023_AI}. During this phase, the technology experiences a peak of inflated expectations, which may explain the weak influence of social factors, such as Social Influence and Power Distance, on its adoption.

\textit{Long Term Orientation} was hypothesized to moderate the relationships between Performance Expectancy and Behavioral Intention, Hedonic Motivation and Behavioral Intention, and Behavioral Intention and Use Behavior. Nevertheless, the moderating effects were not significant. This outcome can be attributed to the immediate and practical focus of software engineering, where decisions are driven more by present benefits and usability rather than long-term cultural values. Engineers prioritize tools that offer immediate performance enhancements and practical benefits, such as LLMs, over considerations influenced by long-term orientation.
Furthermore, the rapid evolution of software development, with engineers at the forefront of adopting new technology, further justifies the absence of any moderating effect. The non-significant influence of hedonic motivation also supports this result.
Additionally, focusing on the performance expectancy construct reveals its crucial impact on the behavioral intention to use LLMs. As LLMs are disruptive and currently perceived as proving themselves as outstanding working companions, the long-term cultural perspectives of practitioners take a secondary role, unable to compete with the immediate benefits provided by such technology~\cite{agossah2023llm, khojah2024beyond}.

\textit{Collectivism vs. Individualism} was hypothesized to moderate the relationships between three independent variables (Performance Expectancy, Effort Expectancy, and Social Influence) with Behavioral Intention, as well as between Behavioral Intention and Use Behavior. However, the moderating effects were not significant. This can largely be explained by the early stage and disruptive nature of LLMs~\cite{Gartner2023_AI}, which are perceived as highly useful~\cite{agossah2023llm, khojah2024beyond}. In such a context, whether individuals lean towards collectivism or individualism may not significantly impact adoption decisions.
This is further supported by the non-significance of Social Influence, a variable closely related to collectivist values. In a field like software engineering, where personal expertise and the practical benefits of tools like LLMs are paramount, the collective social context has less impact on adoption decisions. Engineers are more likely to focus on how LLMs enhance their individual performance and efficiency, even if they value community and teamwork.

\textit{Masculinity vs. Femininity} was hypothesized to moderate the relationship between Performance Expectancy and Behavioral Intention, as well as between Behavioral Intention and Use Behavior. However, the moderating effects were not significant. This finding aligns with broader research suggesting that this cultural dimension does not play a concrete role in technology adoption~\cite{baptista_2015_culture_and_mobile_banking, lin2014investigation}.
In environments characterized by high-performance expectations, both masculine (competitive, achievement-oriented) and feminine (collaborative, quality-of-life-oriented) values converge on the importance of effectiveness and efficiency in tool adoption. Furthermore, the disruptive and innovative nature of LLMs, which offer substantial improvements in productivity and problem-solving capabilities~\cite{agossah2023llm, khojah2024beyond}, likely overshadows any cultural predispositions related to masculinity or femininity.
In summary, the lack of a significant moderating effect of Masculinity vs. Femininity reinforces the idea that practical considerations and the inherent advantages of LLMs are the primary drivers of adoption in the field of software engineering~\cite{russo2024_navigating, khojah2024beyond, agossah2023llm, draxler2023gender}, rather than cultural values related to competitive preferences.

\textit{Uncertainty Avoidance} was hypothesized to moderate the relationships between Performance Expectancy and Behavioral Intention, Social Influence and Behavioral Intention, as well as between Behavioral Intention and Use Behavior. However, the moderating effects were not significant. This result can be explained by several factors specific to the context of software engineering and the nature of LLMs.
In the fast-evolving field of software engineering, practitioners are often accustomed to dealing with high levels of uncertainty and rapid technological changes. This professional culture may reduce the impact of individual differences in Uncertainty Avoidance on technology adoption decisions. Moreover, the perceived significant improvements in efficiency and performance offered by LLMs can mitigate concerns associated with uncertainty~\cite{agossah2023llm, khojah2024beyond}. When the advantages of adopting a disruptive technology are (perceived) evident, the predisposition to avoid uncertainty may have less influence on the decision-making process.

\subsection{Implications for Practise and Researcher}

The study aimed to explore the role of cultural values in the adoption of LLMs in software engineering, guided by the Unified Theory of Acceptance and Use of Technology (UTAUT2) and Hofstede’s cultural dimensions. The results revealed two main points:

\begin{description}[leftmargin=0.3cm]
    \item[\faHandORight] \textbf{Cultural Values are not significant into moderating the adoption of LLMs.} The study found that cultural values do not significantly impact the intention to adopt or the actual adoption of LLMs among software engineers.
    \smallskip
    \item[\faHandORight] \textbf{Traditional adoption factors are outstanding protagonists in explaining the adoption of LLMs.} Within the UTAUT2 framework, several factors were identified as significantly influencing the adoption of LLMs, such as Habit (the tendency to use LLMs as part of regular work activities), Performance Expectancy (the belief that using LLMs will improve job performance), Facilitating Conditions (the availability of resources and support to use LLMs), and Behavioral Intention (the intention to use LLMs in the future).
\end{description}

The implications of these findings are twofold. Firstly, they provide actionable insights for organizations seeking to implement LLMs effectively by highlighting the importance of fostering habitual use, demonstrating performance benefits, ensuring adequate resources and support, and encouraging positive use intentions among practitioners. Secondly, they challenge the assumption that cultural values significantly influence technology adoption in this context, shifting the focus towards other influential factors identified within the UTAUT2 framework.

\subsubsection{Implications for Practitioners}

The factors within the UTAUT2 framework that showed significant impact highlight practical considerations for organizations aiming to encourage LLM adoption.

The primary and most significant implication of our study is the central role that performance expectations play in the intention to adopt LLMs in software engineering. This finding underscores the necessity for organizations aiming to incorporate LLMs into their workflows to demonstrate and monitor the tangible benefits of LLMs in enhancing job performance. Based on our results, we recommend the following actionable insights:

\begin{itemize}
    \item[\faBriefcase] \textbf{Identify, Implement, and Monitor Performance Metrics}: Managers should identify and establish clear performance metrics to measure the impact of LLMs on software engineering tasks. These metrics should be monitored over time to regularly review and communicate ongoing improvements and areas needing optimization to the team. Furthermore, as suggested by the literature~\cite{khojah2024beyond, liu2024empirical}, the importance of specific tasks varies among developers. Therefore, organizations should link the metrics to specific tasks to enhance communication with the team.

    \item[\faBriefcase] \textbf{Adapt and Support Training}: In line with the previous point, organizations should regularly gather feedback from users to understand their experiences and identify any challenges or additional needs. Based on this feedback, organizations should adapt training materials for newcomers to support the adoption of the tool from the onboarding phase.
    
    \item[\faBriefcase] \textbf{Highlight Tangible Benefits}: Related to the previous two points, organizations should create comprehensive reports and presentations that detail improvements in efficiency, accuracy, and overall productivity resulting from LLM adoption. Additionally, as suggested by related literature~\cite{khojah2024beyond, liu2024empirical} and mentioned before, organizations can categorize these improvements by specific tasks (e.g., requirements elicitation, code completion, and decision support) to attract developers' interest based on the areas they find most important.

    \item[\faBriefcase] \textbf{Promote Success Stories}: Organizations should publicize successful projects and individual achievements resulting from the use of LLMs to build enthusiasm and confidence across the organization.

\end{itemize}

Our investigation also highlights the importance of habitual use of LLMs in supporting both the intention to use and the actual use of these tools in the software engineering domain. This result aligns with previous work in the field~\cite{draxler2023gender, agossah2023llm}. Therefore, we recommend the following actionable insights:

\begin{itemize}
    \item[\faBriefcase] \textbf{Encourage Regular Usage}: Organizations and managers should integrate LLM usage into daily workflows to help employees develop a habit. This can be done by facilitating access and integration of these tools in the every day working activity. Nevertheless, organizations should take into consideration ethical concerns. For instance, the Copenhagen Manifesto on the adoption of generative AI in software engineering~\cite{russo2024_copenhagen_manifesto} claims particular attention to the ethical implications and the centric role of human behaviors in such an adoption. Organizations are suggested to consider the manifesto principles while integrating LLMs into their working environment.

    \item[\faBriefcase] \textbf{Provide Continuous Training}: Related to the previous discussion on the performance expectancy, organizations should offer continuous training sessions that reinforce the usage of LLMs over time. Moreover, managers should schedule periodic refresher courses to keep employees updated on new features and best practices for using LLMs. All these activities should take into account the feedback and metrics collected during the development lifecycle on the technology usage.

    \item[\faBriefcase] \textbf{Support a Community of Exchange}: Organizations should establish user communities where employees can share tips, experiences, and support each other in using LLMs. Managers should encourage participation in these communities to foster a culture of collaboration and habitual use.

\end{itemize}

As an ulterior result, facilitating the access to the technology arose as an important factor in the actual adoption of LLMs for software engineering. All the insights proposed so far contribute to a broader effort to facilitate technology adoption by companies. As an ulterior actionable insight, we recommend the following:

\begin{itemize}
    \item[\faBriefcase] \textbf{Invest in building in-house teams}: The scope of the proposed activities necessitates a dedicated workforce. Organizations should identify suitable internal candidates and assemble a team to ensure proper execution and manage the metrics for reporting.
\end{itemize}

To sum up, organizations aiming to implement LLMs effectively should start by integrating them into everyday tasks to develop habitual use. Moreover, it is crucial to highlight the performance benefits by clearly demonstrating how LLMs can enhance productivity and job performance. Furthermore, providing comprehensive support and resources, such as training and technical assistance, will facilitate their use. Additionally, encouraging positive intentions through motivational strategies, incentives, and continuous engagement can foster a positive attitude toward adopting LLMs.

\subsubsection{Implications for Researchers}

This study contributes to the understanding of technology adoption by highlighting that cultural values may not play a significant role in moderating the adoption of LLMs in software engineering. Instead, individual and organizational factors, as outlined in the UTAUT2 framework, are more influential. Researchers could:
\begin{itemize}
    \item[\faLightbulbO] \textbf{Deepening the role of cultural values in technology adoption:} This study challenges the assumption that cultural values play a significant role in the adoption of disruptive technologies like LLMs. Nevertheless, these results can be influenced by a plethora of aspects. Other than conducting longitudinal studies to see if culture starts to be significant (maybe when LLMs will reach the next phases of the Gartner lifecycle), researchers could try to use different scales to measure culture. In this study, we relied on the well-known scale of Yoo because it is largely recognized in the community. Still, another scale like the one Sharma~\cite{sharma2010_cultural_value_scale} exists that measures and represents cultural values with a broader range of constructs~\cite{taras2023_cultural_measures_comparison_individual_level}.
    
    \item[\faLightbulbO] \textbf{Focus on UTAUT2 constructs:} The findings highlight the dominance of UTAUT2 constructs such as Habit, Performance Expectancy, Facilitating Conditions, and Behavioral Intention in explaining the adoption of LLMs. Future research should continue to explore and refine these constructs, considering their practical implications and how they can be leveraged to enhance technology adoption.
    
    \item[\faLightbulbO] \textbf{Conduct longitudinal studies on Habit formation:} The significant role of Habit in LLM adoption indicates a need for longitudinal studies to understand how habitual use of new technologies develops over time. Researchers should explore the processes and interventions that can effectively transform initial use into sustained, routine behavior.
    
    \item[\faLightbulbO] \textbf{Develop frameworks and tools:} Researchers should focus on developing practical frameworks and guidelines based on UTAUT2 constructs to assist organizations in implementing new technologies. These frameworks should provide actionable strategies for fostering habitual use, demonstrating performance benefits, ensuring adequate resources, and encouraging positive use intentions. Moreover, researchers could take the challenge of developing a recommendation system to support managerial figures in the process of integrating LLMs into the development workflow without damaging the work environment.
\end{itemize}

By addressing these implications, the research community can advance the understanding of technology adoption in specialized fields and contribute to more effective implementation strategies for emerging technologies.

\subsection{Threats to Validity and Limitations}
\label{sec_limitation} 

As previously mentioned, our study primarily involved quantitative investigations supported by statistical analysis. Consequently, we discuss the threats to validity by examining statistical conclusion validity, internal validity, construct validity, and external validity as defined by Wohlin et al.~\cite{wohlin2012_experimentation}.

\textbf{Conclusion Validity:} This type of validity pertains to limitations that affect the ability to draw conclusions about the relationships between the independent and dependent variables~\cite{wohlin2012_experimentation}. Threats in this category are mainly associated with the tests employed for the analysis. To mitigate this threat, we heavily relied on PLS-SEM statistics, known for their robustness across various scenarios. We meticulously followed the process described by Hair et al.~\cite{hair_2014_PLS} in their comprehensive work on using PLS-SEM. Additionally, we used SmartPLS, a state-of-the-art software utilized in over 1,000 peer-reviewed articles~\cite{hair_2014_PLS}.

\textbf{Internal Validity:} Internal validity concerns the risk that the investigation did not account for factors influencing the dependent variable~\cite{wohlin2012_experimentation}. To address this, we relied on solid and well-validated theories. Given our focus on technology adoption, we utilized the Unified Theory of Acceptance and Use of Technology (UTAUT2), which is well-suited for investigating similar phenomena~\cite{Venkatesh_2012_UTAUT2}. For studying cultural values, we employed the Hofstede cultural framework~\cite{hofstede_1980_cultural_model_1, hofstede_2011_cultural_model_2}, a widely adopted instrument operationalized for this purpose. Furthermore, we applied a set of filters to our participant sample to ensure it represented the target population while maintaining heterogeneity.

\textbf{Construct Validity:} Construct validity addresses the measurement and tools used to represent the study variables~\cite{wohlin2012_experimentation}. All variables in our theoretical model were measured using validated instruments~\cite{Venkatesh_2012_UTAUT2, yoo_2011_CVSCALE_measuring_culture}. Specifically, for measuring culture, we avoided the common error of using country of origin combined with Hofstede's values. Instead, we relied on scales designed to measure culture at the individual level~\cite{yoo_2011_CVSCALE_measuring_culture}. We designed the questionnaires following the most recent guidelines in the field and employed various strategies, such as question randomization and attention-check questions, to enhance the reliability of the results~\cite{kitchenham2008_PersonalOpinionSurveys, ralph_2020_empirical_standards, danilova2021_developers_questions}.

\textbf{External Validity:} External validity concerns the generalizability of the findings to the entire population~\cite{wohlin2012_experimentation}. We filtered the entire Prolific population to ensure that participants met the characteristics ideal for our investigation. Additionally, we collected a sufficient number of data according to G*Power recommendations~\cite{faul_2009_GPower}. However, generalizability is a complex issue. Given that the technology is still in its early stages~\cite{Gartner2023_AI}, we anticipate that different results might emerge in the future.

\steSummaryBox{\faHandORight \hspace{0.05cm} UTAUT2 and Cultural Values for LLMs in Software Engineering: Summary of the discussion.}{This study investigated the impact of cultural values as moderator in the adoption of Large Language Models (LLMs) in software engineering using the Unified Theory of Acceptance and Use of Technology (UTAUT2) framework. The research revealed that cultural values did not significantly moderate the intention to adopt or the actual adoption of LLMs. Instead, the UTAUT2 constructs—habit, performance expectancy, facilitating conditions, and behavioral intention—were the key factors affecting adoption. Performance expectancy was the most influential, emphasizing the importance of perceived benefits in adopting new technologies. Habit also played a crucial role, indicating that once LLMs become part of routine work, their continued use is likely. Facilitating conditions, such as resources and support, were critical for actual use, while Behavioral Intention directly influenced adoption.
For practitioners, the study suggests focusing on integrating LLMs into daily workflows, showcasing their performance benefits, ensuring robust support and resources, and fostering positive use intentions through training and awareness programs. For researchers, the findings highlight the need to explore UTAUT2 constructs further, particularly habit formation, and to develop practical frameworks for technology adoption. Additionally, the study challenges the assumption that cultural values significantly impact technology adoption in this context, suggesting a shift in focus toward individual and organizational factors.}

\section{Conclusions}
\label{sec_conclusion}

The objective of this study was to explore the adoption of Large Language Models (LLMs) in software development, with a focus on the potential moderating effects of practitioners' cultural values. We concentrated first on typical factors influencing technology adoption, as derived from the Unified Theory of Acceptance and Use of Technology (UTAUT2), and second on cultural values as defined by Hofstede.

Our statistical analysis, conducted using Partial Least Squares Structural Equation Modeling (PLS-SEM), revealed that habit and perceived performance are the primary drivers of integrating LLMs into daily routines. In contrast, the moderating effect of cultural values was not significant. This outcome can be attributed to the complexity of software development, which is heavily influenced by evolving socio-technical factors and the increasing demand for rapid release cycles. Furthermore, the disruptive and highly beneficial nature of LLMs likely overshadows cultural influences, providing a clearer justification for our findings. In addition, our results align with what was identified by the Human-AI Collaboration and Adaptation Framework (HACAF)~\cite{russo2024_navigating}.

For practitioners, our findings have direct and practical implications. They suggest integrating LLMs into daily workflows, clearly demonstrating their performance benefits, providing comprehensive support and resources, and fostering positive use intentions through motivational strategies. The findings also underscore the need for further research to explore cultural values, conduct longitudinal studies on habit formation, and develop practical frameworks and tools to assist organizations in effectively implementing new technologies like LLMs.

For our future research agenda, we plan to explore additional ways to integrate cultural values into the technology adoption of LLMs in software engineering, beyond using the extended UTAUT2 framework, which our research has validated. Additionally, we aim to investigate alternative scales to the CVSCALE proposed by Yoo et al.~\cite{yoo_2011_CVSCALE_measuring_culture} to measure cultural values, as differences, though minimal, can exist among various representations~\cite{taras2023_cultural_measures_comparison_individual_level}. Finally, we advocate for longitudinal studies to understand how the behavioral intention of software engineers evolves as the technology matures and moves beyond the early adoption stage~\cite{Gartner2023_AI}.


\begin{acks}
The authors thank the questionnaire participants for their valuable responses and for providing the essential data required to conduct this research. This work has been partially supported by the European Union through the Italian Ministry of University and Research, Project PRIN 2022 ``\textsl{QUAL-AI}: Continuous Quality Improvement of AI-based Systems'' (grant n. 2022B3BP5S, CUP: H53D23003510006). 
\end{acks}

\bibliographystyle{ACM-Reference-Format}
\bibliography{bibliography}

\clearpage
\appendix

\section{Appendix A — Questionnaires Items}
\label{AppendixA_label}

\begin{longtable}[H]{p{0.2\linewidth} p{0.1\linewidth} p{0.55\linewidth} p{0.05\linewidth}}
    \caption{Items Description. The items prefixed with an asterisk (*) were dropped because of their insufficient loading onto their latent variable.}\label{table_appendix_ind}\\\hline
    \endfirsthead
    \caption* {Table \ref{table_appendix_ind} Continued. The items prefixed with an asterisk (*) were dropped because of their insufficient loading onto their latent variable.}\\\hline
    \endhead
    \endfoot
    \hline
    \endlastfoot
    
        \textbf{Constructs} & \textbf{Item ID} & \textbf{Questions} & \textbf{Ref.}\\ \hline
        \multirow{5}{*}{Power Distance Index} & PD\_1 & People in higher positions should make most decisions without consulting people in lower positions. & \cite{yoo_2011_CVSCALE_measuring_culture}\\
        & PD\_2 & People in higher positions should not ask the opinions of people in lower positions too frequently. & \cite{yoo_2011_CVSCALE_measuring_culture}\\
        & PD\_3 & (*) People in higher positions should avoid social interaction with people in lower positions. & \cite{yoo_2011_CVSCALE_measuring_culture}\\
        & PD\_4 & People in lower positions should not disagree with decisions by people in higher positions. & \cite{yoo_2011_CVSCALE_measuring_culture}\\
        & PD\_5 & (*) People in higher positions should not delegate important tasks to people in lower positions. & \cite{yoo_2011_CVSCALE_measuring_culture}\\
        \hline
        \multirow{5}{*}{Uncertainty Avoidance} & UN\_1 & (*) It is important to have instructions spelled out in detail so that I always know what I’m expected to do. & \cite{yoo_2011_CVSCALE_measuring_culture}\\
        & UN\_2 & It is important to closely follow instructions and procedures. & \cite{yoo_2011_CVSCALE_measuring_culture}\\
        & UN\_3 & (*) Rules and regulations are important because they inform me of what is expected of me. & \cite{yoo_2011_CVSCALE_measuring_culture}\\
        & UN\_4 & (*) Standardized work procedures are helpful. & \cite{yoo_2011_CVSCALE_measuring_culture}\\
        & UN\_5 & Instructions for operations are important. & \cite{yoo_2011_CVSCALE_measuring_culture}\\
        \hline
        \multirow{6}{*}{Collectivism} & CO\_1 & (*) Individuals should sacrifice self-interest for the group. & \cite{yoo_2011_CVSCALE_measuring_culture}\\
        & CO\_2 & Individuals should stick with the group even through difficulties. & \cite{yoo_2011_CVSCALE_measuring_culture}\\
        & CO\_3 & Group welfare is more important than individual rewards. & \cite{yoo_2011_CVSCALE_measuring_culture}\\
        & CO\_4 & Group success is more important than individual success. & \cite{yoo_2011_CVSCALE_measuring_culture}\\
        & CO\_5 & (*) Individuals should only pursue their goals after considering the welfare of the group. & \cite{yoo_2011_CVSCALE_measuring_culture}\\
        & CO\_6 & Group loyalty should be encouraged even if individual goals suffer. & \cite{yoo_2011_CVSCALE_measuring_culture}\\
        \hline
        \multirow{4}{*}{Masculinity} & MA\_1 & It is more important for men to have a professional career than it is for women. & \cite{yoo_2011_CVSCALE_measuring_culture}\\
        & MA\_2 & Men usually solve problems with logical analysis; women usually solve problems with intuition. & \cite{yoo_2011_CVSCALE_measuring_culture}\\
        & MA\_3 & Solving difficult problems usually requires an active, forcible approach, which is typical of men. & \cite{yoo_2011_CVSCALE_measuring_culture}\\
        & MA\_4 & (*) There are some jobs that a man can always do better than a woman. & \cite{yoo_2011_CVSCALE_measuring_culture}\\
        \hline
        \multirow{6}{*}{Long Term Orientation} & LT\_1 & (*) Careful management of money (Thrift) & \cite{yoo_2011_CVSCALE_measuring_culture}\\
        & LT\_2 & Going on resolutely in spite of opposition (Persistence) & \cite{yoo_2011_CVSCALE_measuring_culture}\\
        & LT\_3 & (*) Personal steadiness and stability & \cite{yoo_2011_CVSCALE_measuring_culture}\\
        & LT\_4 & Long-term planning & \cite{yoo_2011_CVSCALE_measuring_culture}\\
        & LT\_5 & (*) Giving up today’s fun for success in the future & \cite{yoo_2011_CVSCALE_measuring_culture}\\
        & LT\_6 & (*) Working hard for success in the future & \cite{yoo_2011_CVSCALE_measuring_culture}\\
        \hline
        \multirow{5}{*}{Performance Expectancy} & PE\_1 & I find the use of LLM useful in supporting my job & \cite{Venkatesh_2012_UTAUT2}\\
        & PE\_2 & Using LLMs to support my job increases my productivity & \cite{Venkatesh_2012_UTAUT2}\\
        & PE\_3 & Using LLMs to support my job helps me accomplish things more quickly & \cite{Venkatesh_2012_UTAUT2}\\
        & PE\_4 & Using LLMs to support my job increases my chances of achieving things that are important to me & \cite{Venkatesh_2012_UTAUT2}\\
        & PE\_5 & Using LLMs would improve the quality of work I do & \cite{Venkatesh_2012_UTAUT2}\\
        \hline
        \multirow{5}{*}{Performance Expectancy} & PE\_1 & I find the use of LLM useful in supporting my job & \cite{Venkatesh_2012_UTAUT2}\\
        & PE\_2 & Using LLMs to support my job increases my productivity & \cite{Venkatesh_2012_UTAUT2}\\
        & PE\_3 & Using LLMs to support my job helps me accomplish things more quickly & \cite{Venkatesh_2012_UTAUT2}\\
        & PE\_4 & Using LLMs to support my job increases my chances of achieving things that are important to me & \cite{Venkatesh_2012_UTAUT2}\\
        & PE\_5 & Using LLMs would improve the quality of work I do & \cite{Venkatesh_2012_UTAUT2}\\
        \hline
        \multirow{6}{*}{Effort Expectancy} & EE\_1 & Learning how to use LLMs to support my job is easy for me & \cite{Venkatesh_2012_UTAUT2}\\
        & EE\_2 & My interaction with LLMs to support my job is clear and understandable & \cite{Venkatesh_2012_UTAUT2}\\
        & EE\_3 & I find LLMs easy to use in supporting my job & \cite{Venkatesh_2012_UTAUT2}\\
        & EE\_4 & It is easy for me to become skillful at using LLMs to support my job & \cite{Venkatesh_2012_UTAUT2}\\
        & EE\_5 & I would find LLMs flexible to interact with during my job & \cite{Venkatesh_2012_UTAUT2}\\
        & EE\_6 & Using LLMs would make it easier to do my job & \cite{Venkatesh_2012_UTAUT2}\\
        \hline
        \multirow{5}{*}{Social Influence} & SI\_1 & People who are important to me think that I should use LLMs to support my job & \cite{Venkatesh_2012_UTAUT2}\\
        & SI\_2 & People who influence my behavior think that I should use LLMs to support my job & \cite{Venkatesh_2012_UTAUT2}\\
        & SI\_3 & (*) LLMs use is a status symbol in my job environment & \cite{Venkatesh_2012_UTAUT2}\\
        & SI\_4 & (*) I use LLMs because of the proportion of coworkers who use the system & \cite{Venkatesh_2012_UTAUT2}\\
        & SI\_5 & (*) People in my organisation who use LLMs have more prestige than those who do not & \cite{Venkatesh_2012_UTAUT2}\\
        \hline
        \multirow{4}{*}{Facilitating Condition} & FC\_1 & I have the resources necessary to use LLMs to support my job & \cite{Venkatesh_2012_UTAUT2}\\
        & FC\_2 & I have the knowledge necessary to use LLMs to support my job & \cite{Venkatesh_2012_UTAUT2}\\
        & FC\_3 & LLMs are compatible with other technologies I use in my job & \cite{Venkatesh_2012_UTAUT2}\\
        & FC\_4 & (*) I can get help from others when I have difficulties using LLMs to support my job & \cite{Venkatesh_2012_UTAUT2}\\
        \hline
        \multirow{3}{*}{Hedonic Motivation} & HM\_1 & Using LLMs to support my job is fun & \cite{Venkatesh_2012_UTAUT2}\\
        & HM\_2 & Using LLMs to support my job is enjoyable & \cite{Venkatesh_2012_UTAUT2}\\
        & HM\_3 & Using LLMs to support my job is entertaining & \cite{Venkatesh_2012_UTAUT2}\\
        \hline
        \multirow{4}{*}{Habit} & HB\_1 & The use of LLMs to support my job has become a habit for me & \cite{Venkatesh_2012_UTAUT2}\\
        & HB\_2 & I am addicted to using LLMs to support my job & \cite{Venkatesh_2012_UTAUT2}\\
        & HB\_3 & I must use LLMs to support my job & \cite{Venkatesh_2012_UTAUT2}\\
        & HB\_4 & Using LLMs to support my job has become natural to me & \cite{Venkatesh_2012_UTAUT2}\\
        \hline
        \multirow{3}{*}{Price Value} & PV\_1 & LLMs that I use to support my job are reasonably priced & \cite{Venkatesh_2012_UTAUT2}\\
        & PV\_2 & LLMs that I use to support my job are a good value for the money & \cite{Venkatesh_2012_UTAUT2}\\
        & PV\_3 & At the current price, LLMs provide a good value & \cite{Venkatesh_2012_UTAUT2}\\
        \hline
        \multirow{3}{*}{Behavioral Intention} & BI\_1 & I intend to continue using LLMs to support my job in the future & \cite{Venkatesh_2012_UTAUT2}\\
        & BI\_2 & I will always try to use LLMs to support my job & \cite{Venkatesh_2012_UTAUT2}\\
        & BI\_3 & I plan to continue to use LLMs to support my job frequently & \cite{Venkatesh_2012_UTAUT2}
\end{longtable}

\end{document}